\documentclass[11pt]{article}

\usepackage{amsmath}
\usepackage{amssymb}
\usepackage[font=small,labelfont=bf]{caption}

\usepackage{graphicx}
\usepackage{float}

\usepackage{cite}

\usepackage{color}

\usepackage[labelfont=bf,labelsep=period,justification=raggedright]{caption}

\makeatletter
\renewcommand{\@biblabel}[1]{\quad#1.}
\makeatother

\date{}

\definecolor{newtextcolor1}{cmyk}{0,1,1,0} 
\definecolor{newtextcolor2}{cmyk}{0.01,0.7,1,0} 
\definecolor{newtextcolor3}{cmyk}{1,1,0.01,0} 
\definecolor{newtextcolor4}{cmyk}{1,0.01,1,0} 
\definecolor{newtextcolor5}{cmyk}{1.,0.,0,0} 

\begin{document}

\begin{flushleft}
{\Large
\textbf{Astrocytic Ion Dynamics: Implications for Potassium Buffering and Liquid Flow}
}
\\
Geir Halnes$^{1,\ast}$,
Klas H. Pettersen$^{2}$,
Leiv \O yehaug$^{3}$,
Marie E. Rognes$^{4,5}$,
Gaute T. Einevoll$^{1,6}$
\\
\bf{1} Faculty of Science and Technology, Norwegian University of Life Sciences, {\AA}s, Norway
\\
\bf{2} Letten Centre and GliaLab, Centre for Molecular Medicine, University of Oslo, Oslo, Norway
\\
\bf{3} Faculty of Technology, Art and Design, OsloMet - Oslo Metropolitan University, Oslo, Norway
\\
\bf{4} Simula Research Laboratory, Fornebu, Norway
\\
\bf{5} Dept. of Mathematics, University of Oslo, Oslo, Norway
\\
\bf{6} Dept. of Physics, University of Oslo, Oslo, Norway

$\ast$ E-mail: geir.halnes@nmbu.no
\end{flushleft}

\section*{Abstract} 
We review modeling of astrocyte ion dynamics with a specific focus on the implications of so-called spatial potassium buffering where excess potassium in the extracellular space (ECS) is transported away to prevent pathological neural spiking. The recently introduced
Kirchoff-Nernst-Planck (KNP) scheme for modeling ion dynamics in astrocytes (and
brain tissue in general) is outlined and used to study such spatial buffering.
We next describe how the ion dynamics of astrocytes may regulate microscopic
liquid flow by osmotic effects and how such microscopic flow can be linked to
whole-brain macroscopic flow. The chapter thus describes key elements in a
putative multiscale theory with astrocytes linking neural activity on a
microscopic scale to macroscopic fluid flow.

{\bf Keywords:} Tissue modelling; Ion concentration dynamics; Electrodiffusion; Neuron-glia interactions; Potassium buffering.


\section{Introduction} \label{sec:intro}
Brain function is fundamentally about movement of ions and molecules. When
modelling neuronal dynamics, it is common to neglect the dynamics of individual
ion species, and rather just simulate the net electrical currents and resulting
changes in membrane potentials. The justification is that on the typical time
scale of neural integration (milliseconds), the concentrations of the major
charge carriers in brain tissue (i.e., Na$^+$, K$^+$, Cl$^-$, ...) vary little.
For astrocytic dynamics the situation is different. For one, the typical time
scale for astrocytic membrane processes are longer, i.e., seconds rather than
milliseconds, so that the fixed ion-concentration assumption is \emph{a priori}
more dubious. Further, many key astrocytic functions are related just to their responses
to shifts in extracellular ion concentrations  \cite{Wang2008}.

Astrocytes have several homeostatic functions in the brain. They provide
metabolic support for neurons, synthesize extracellular matrix proteins,
adhesion molecules, and trophic factors controlling neuronal maturation, are
involved in the formation of blood vessels and in maintenance of the blood-brain
barrier, and maintain the ECS via uptake of K$^+$ and neurotransmitters
\cite{Wang2008}. In the present chapter we describe theoretical frameworks for
modelling astrocytic ion concentration dynamics, with particular focus on K$^+$
clearance mechanisms, and possible consequences for micropscopic liquid flow in
the brain.

When neurons fire action potentials (APs), they absorb Na\textsuperscript{+} and
expel K\textsuperscript{+} from/to the ECS. At low to moderate firing
frequencies, the time interval between two APs is sufficient for neuronal
Na\textsuperscript{+}\textendash{}K\textsuperscript{+} pumps to restore the
baseline levels of Na\textsuperscript{+} and K\textsuperscript{+}. Ion
concentrations then remain essentially constant over time
\cite{somjen_ions_2004}. However, during periods of intense neural signalling,
the neuronal ion pumps may fail to keep up, and ECS ion concentrations can
change significantly \cite{frankenhaeuser_after-effects_1956, Cordingley1978,
Dietzel1989, Gardner-Medwin1983, Chen2000, Haj-Yasein2014}. The most critical
effects of this relates to changes in the K\textsuperscript{+}-concentration in
the ECS, which can increase from a relatively low baseline level of around 3 mM
up to levels between 8 and 12 mM during non-pathological conditions
\cite{Hertz2013, Chen2000, Newman1993}. Increases beyond this can occur under
pathological conditions such as hypoxia, anoxia, ischemia epilepsy and spreading
depression \cite{Sykova2008, Enger2015, Park2006, florence_role_2009}. During
spreading depression, K\textsuperscript{+}-concentrations in the ECS can become
as high as 60 mM \cite{somjen2001mechanisms}. 

Astrocytes have several membrane mechanisms for local uptake of excess
K\textsuperscript{+} \cite{Orkand1966, lux_ionic_1986,somjen_ions_2004,
Gardner-Medwin1983, Newman1993, Chen2000, Wang2008, oyehaug_dependence_2011}. They are therefore likely to play a role in pathological conditions related to ion dynamics in the
ECS, and evidence suggest that changes in astrocytic function is implicated in
spreading depression, epilepsy and ischemia \cite{de2008dysfunctional, Nedergaard2005}. 
In Section \ref{sec:pathology} we briefly review some modelling works that
have investigated the relationship between astrocytic regulation of ECS
K\textsuperscript{+} and neuronal firing patterns.

\begin{figure}[!ht]
\begin{center}
\includegraphics[width=4in]{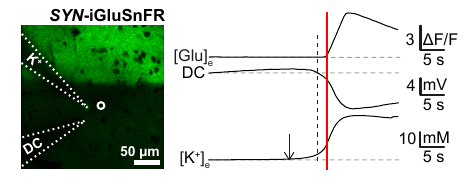}
\end{center}
\caption{
{\bf Changes in extracellular K\textsuperscript{+} concentration, extracellular
glutamate and extracellular potential (DC) during cortical spreading
depression.} 
\emph{Left:} The green fluorescent tracer shows the extracellular
glutamate wave, which travels with a speed of $50~\mathrm{\mu m/s}$. The dashed
white lines indicate the positioning of the recording electrodes, while the
solid white line shows the region of interest (ROI) for glutamate. \emph{Right:}
The glutamate signal within the ROI ([$Glu^{+}$]$_e$, upper), extracellular
potential (DC, middle) and extracellular K\textsuperscript{+} concentration
([$K^{+}$]$_e$, lower), all as functions of time. The figure is modified from
Fig.~4 in Enger et al.~\cite{Enger2015}. }
\label{Fdata}
\end{figure}

Increases in the ECS K\textsuperscript{+}-concentration are often accompanied by
a slow negative shift in the ECS potential. These potential shifts can be on the
order of a few millivolts \cite{Sykova2008, Kriv1975, Lothman1975, Dietzel1989,
Cordingley1978}. Figure \ref{Fdata} shows an extreme example from an experiment
on cortical spreading depression, where a K\textsuperscript{+}-concentration
shift of 20-30 mM was accompanied by a 5-10 mV shift in the ECS potential
\cite{Enger2015}. It has been argued that this link between sustained potentials
and the ECS K\textsuperscript{+} concentration is the signature of glial
K\textsuperscript{+} buffering currents \cite{Dietzel1989}. In addition to local
uptake and storage in astrocytes \cite{Dietzel1989, Coles1986}, clearance of
excess K\textsuperscript{+} from local high-concentration regions may occur by
transport through the ECS \cite{Nicholson2000}, and by a process coined
\emph{spatial K\textsuperscript{+} buffering}, where astrocytes take up
K\textsuperscript{+} from high concentration regions, transport it
intracellularly, and release it in regions where the ECS concentration is lower
\cite{Orkand1966, Gardner-Medwin1983, Newman1993, Chen2000, Halnes2013}. As
astrocytes often are interconnected by gap junctions into a syncytium,
intracellular transport can in principle occur over quite large distances and
via several cells \cite{Chen2000}. The relative importance of these different
clearance mechanisms are under debate \cite{Macaulay2012}. As these processes
involves variations in both ionic concentrations and electrical potentials,
ionic transports may be propelled both by voltage- and concentration gradients
\cite{Kofuji2004}. In Section \ref{sec:Kbuff}, we explore spatial
K\textsuperscript{+} buffering in detail, and present a recently developed
biophysical modelling formalism for simulating electrodiffusive processes - the
\emph{Kirchhoff-Nernst-Planck (KNP)} scheme.

Changes in ionic concentrations will also induce osmotic pressures, which may
lead to water uptake, astrocyte swelling and micropscopic liquid flow in the
brain ~\cite{ostby_astrocytic_2009, oyehaug_dependence_2011}. A complete modelling framework that
combines spatial K\textsuperscript{+} buffering processes with processes related
to water flows is currently lacking, but in Section
\ref{sec:Osmosis_and_astrocyte_swelling}, we review the theoretical foundation
that such a framework would need to be based on.

\section{Influence of glial $\mathrm{K}^{+}$ buffering on neuronal activity} \label{sec:pathology}
\begin{figure}
\centering
\includegraphics[width=3in]{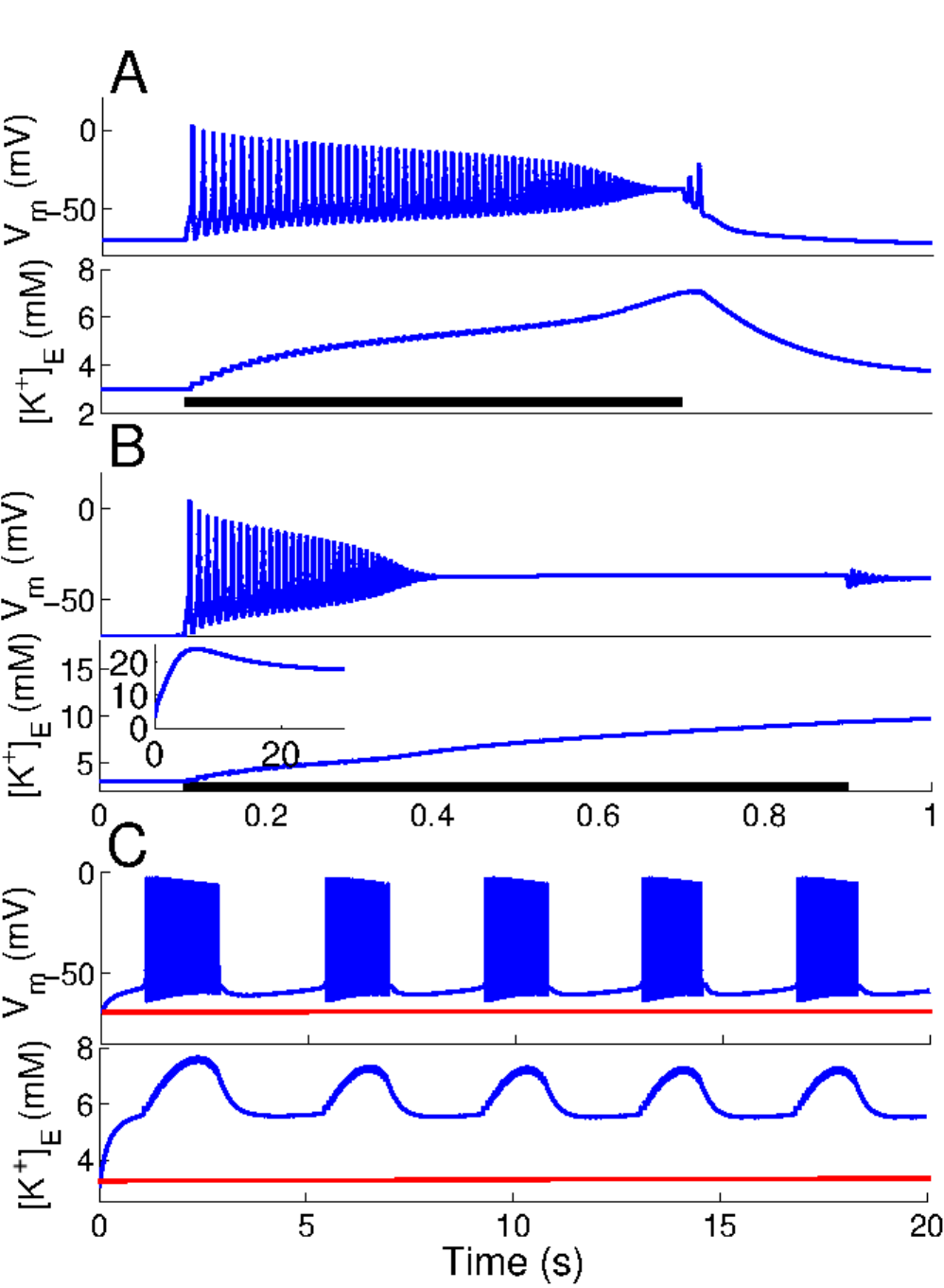}
\caption{\textbf{Dynamical repertoire of neuron-glia model.}
The model \cite{oyehaug_dependence_2011} included voltage-gated sodium and potassium
channels and the NaKATPase pump in the neuronal
membrane (following Kager et al.~\cite{kager_simulated_2000}),
and sodium, potassium and chloride channels, as well as the NaKATPase
pump, the NKCC1 and NBC cotransporters and water channels
in the glial membrane \cite{ostby_astrocytic_2009}.
To ensure charge neutrality
of the ECS, the assumption was made that the amounts of sodium and
potassium crossing the neuronal membrane were identical so that the
cation current was exactly zero. A, B: Dynamics of neuronal membrane potential $V_{m}$ (top)
and $\left[K^{+}\right]_E$ (bottom) when the neuron was stimulated
(A) with a short pulse of relatively small magnitude, and (B) with
a longer pulse of larger magnitude (inset in (B) shows long-time dynamics).
C: Dynamics of neuronal membrane potential $V_{m}$ (top)
and $\left[K^{+}\right]_{E}$ (bottom) compared to
resting conditions (red lines) in absence
of stimulation with a reduction of the maximum rate of the glial
sodium-potassium pump by a factor $0.62$, see \cite{oyehaug_dependence_2011} for details.
\label{fig:Dynamical-repertoire}}
\end{figure}

The effect of ion-concentration shifts on neuronal activity can to a large
extent be explained by its impact on the reversal potential $e_k$ of an ion
species $k$:
\begin{equation}
e_k =  \frac{\psi }{z_k} \log([k]_{E}/{[k]_{I}}).
\label{ek}
\end{equation}
As Eq. \ref{ek} shows, the reversal potential depends on the ion concentrations
on the outside ($\left[k\right]_E$) and inside ($\left[k\right]_I$) of the
membrane ($z_k$ ia the valence of ion species $k$, and $\psi = RT/F$ where $R$
is the gas constant, $T$ the absolute temperature, and $F$ is Faraday's
constant). For example, when $\mathrm{K}^{+}$ accumulates in the ECS, the
$\mathrm{K}^{+}$ reversal potential will increase. As a consequence, neurons
will become depolarized and brought closer to their firing threshold, which in
turn may lead to enhanced neuronal activity and to further increases in
$\left[K^{+}\right]_{e}$.
Computational models that study disorders related to extracellular
$\mathrm{K}^{+}$ are many (see e.g., \cite{oyehaug_dependence_2011,
ullah_models_2009, hubel_dynamics_2014,
kager_simulated_2000,kager_seizure-like_2006,cressman_influence_2009,
florence_role_2009, Park2006, Somjen2008, sibille2015neuroglial}). 
Figure ~\ref{fig:Dynamical-repertoire} shows an
example from a previous modelling study where we explored the positive feedback
loop described above \cite{oyehaug_dependence_2011}. The model used in
\cite{oyehaug_dependence_2011} combined a Hodgkin-Huxley type neuron model
\cite{kager_simulated_2000} with a detailed model of glial membrane ion and
water transport \cite{ostby_astrocytic_2009}.

Figs.~\ref{fig:Dynamical-repertoire}A,B show the results of two numerical
simulations of the neuron-glia model when the neuron was stimulated by
electrical currents of different durations and magnitudes. Elevated levels of
$\left[K^{+}\right]_E$ were observed both with the briefest current stimulation
(Fig.~\ref{fig:Dynamical-repertoire}A) and a longer and stronger current
stimulation (Fig.~\ref{fig:Dynamical-repertoire}B). However, in terms of the
resulting spiking patterns, the difference was striking: with the briefest
stimulation, the neuron fired several APs after the stimulation had ceased, and
then returned to the resting state. By contrast, the larger and longer stimulus
drove the neuron into a deactivated state where the neuron lingered at a
depolarized level with a non-zero K\textsuperscript{+}-efflux. In this case,
$\left[K^{+}\right]_{E}$ reached very high levels
(Fig.~\ref{fig:Dynamical-repertoire}B, inset). The behaviour encountered in the
former situation is referred to as \emph{spontaneous discharge}, while the
latter situation is referred to as \emph{depolarization block}.

Whereas Figs.~\ref{fig:Dynamical-repertoire}A and B illustrate the interplay
between neuronal activity and $\left[K^{+}\right]_{E}$,
Fig.~\ref{fig:Dynamical-repertoire}C shows more directly the effect that
astrocytes may have on neurodynamics. Here, the neuron received no external
input. When the astrocyte had the default parametrization,
$\left[K^{+}\right]_{E}$ was maintained at a low level, and the neuron was at
rest (red lines). However, when we modelled astrocytic dysfunction as a
reduction in the astrocyte's Na\textsuperscript{+}/K\textsuperscript{+}-pump
rate, $\left[K^{+}\right]_{E}$ increased, and drove the neuron into periodic
bursting. This bursting behaviour, also encountered in another model of
neuron-glia interaction \cite{cressman_influence_2009}, was previously observed
in experiments and has been interpreted as relevant to some types of epilepsy
\cite{jensen_role_1997,ziburkus_interneuron_2006}. Further, in human epilepsy
patients it has been found that the overall activity of NaKATPase is reduced
\cite{grisar_contribution_1992}, consistent with the results in
Fig.~\ref{fig:Dynamical-repertoire}, showing that neurons are more likely to
fire spontaneous discharges, go into depolarization block or display bursting
behaviour when the glial sodium-potassium pump rate was reduced.

Fig.~\ref{fig:Dynamical-repertoire} just showed a few illustrative examples on
how K\textsuperscript{+} buffering can be relevant for neurodynamics. For a more
thorough analysis of this, we refer to the original publication
\cite{oyehaug_dependence_2011}, where we presented a bifurcation analysis that
systematically mapped out how different neuronal firing states depended on
variations in selected model parameters. In the following, we direct our focus
towards the main topic of this book chapter, i.e., on the glial
K\textsuperscript{+} buffering process as such.

\section{An electrodiffusive model of spatial K\textsuperscript{+} buffering by astrocytes} \label{sec:Kbuff}
Unlike neurons, glial cells are not predominantly driven by synaptic input, and
do not produce fast unitary events such as action potentials. To a large extent,
astrocytic membrane dynamics appears rather to be driven by slow variations in
ECS ion concentrations and voltage differences between the ECS and the
intracellular astrocytic space. Astrocyte resting membrane potentials have been
reported to be heterogeneous, in part because of the heterogeneity in
K\textsuperscript{+} conductance \cite{verkhratsky2013glial,
zhang2010astrocyte}. However, as a tentative approximation, we assume that all
astrocytes within a (relatively large) spatial region receive roughly the same
input and perform roughly the same processing, and can be collapsed into a
single astrocyte domain, representing the mass average astrocytic response (this
assumption was motivated previously \cite{Gardner-Medwin1983, Chen2000,
Halnes2013}). Furthermore, astrocytes are often coupled with gap junctions,
allowing intracellular ionic transports over larger distances through a
syncytium of many interconnected cells \cite{Chen2000, Enger2015}. Intracellular
transports in the astrocyte domain are therefore conceptually similar to ECS
transports, and not limited by the spatial extension of a single cell.

\begin{figure}[!ht]
\begin{center}
\includegraphics[width=7in]{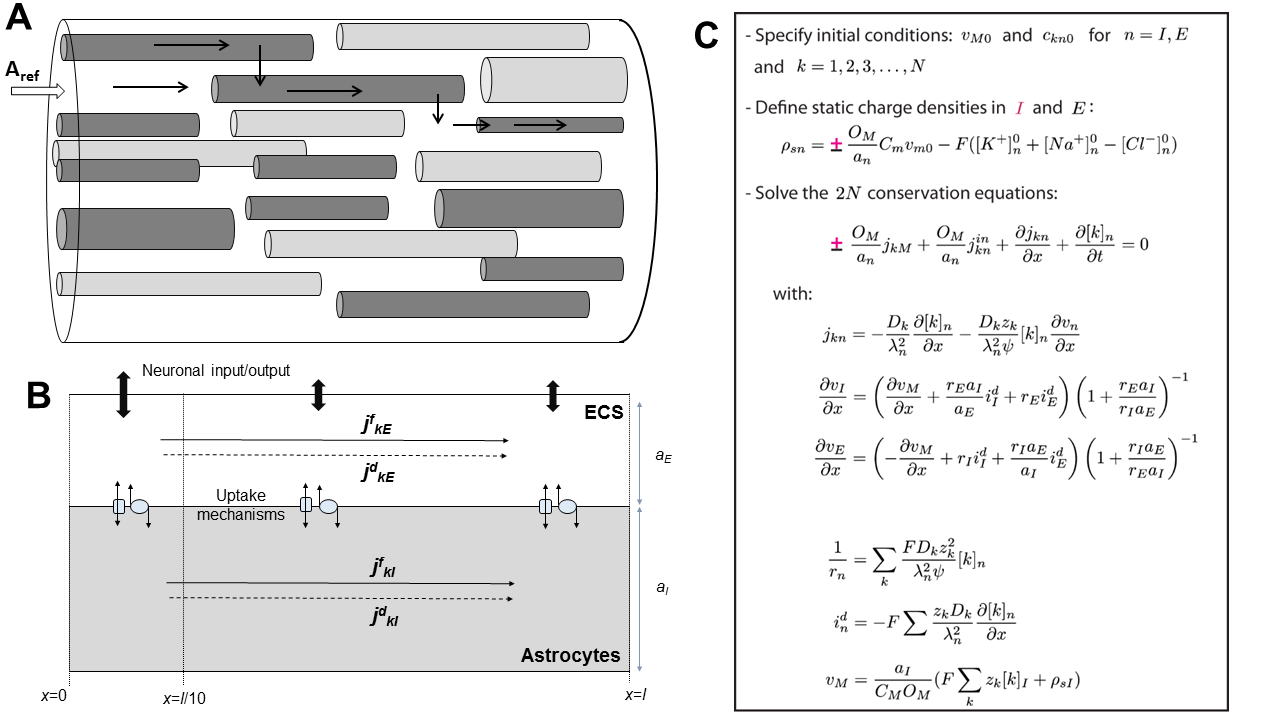}
\end{center}
\caption{
{\bf A two domain-model for ion concentration dynamics in astrocytes and
extracellular space when macroscopic transport is essentially one-dimensional.}
(A) A piece of brain tissue with cross section area $A_{\rm ref}$ and an
arbitrary extension $l$ in the $x$-direction. Tissue astrocytes (dark grey)
participate in the transport process, other cells do not (light grey). (B) The
interior of all astrocytes represented as a single domain (essentially an
equivalent cylindrical cable coated by ECS), where all parts of all astrocytes
at a given location $x$ are assumed to undergo the same activity. Here, $a_I$
and $a_E$ are, respectively, the fractions of $A_{\rm ref}$ occupied by
astrocytes and the ECS. Due to the presence of other cells (non-participatory),
we generally have that $a_I+a_E<1$. The astrocyte domain exchanges ions with the
ECS via astrocytic membrane mechanisms. Axial transport processes in the ECS and
inside astrocytes occur due to electrical migration ($j^f$) and diffusion ($j^d$). Neuronal
activity is represented as an external input to the ECS. Enhanced neuronal input
to a specific region of the ECS (e.g., the region $0<x<l/10$ in the figure) will
evoke spatial buffering processes in the system. (C) A set of equations that
summarizes the electrodiffusive formalism. In equations containing the symbol
"$\pm$", "$+$" should be used for intracellular domain ($n=I$) and "$-$" should
be used for the extracellular domain ($n=E$). The formalism is general to the
choice of $j_{kM}$, representing system specific membrane mechanisms (ion pumps,
ion channels, cotransporters ect.), which will depend on the case at hand.
External input to the system must also be specified. The figure was modified
from \cite{Halnes2013}. }
\label{Fgeo}
\end{figure}

A framework for modelling long term astrocytic processing on an extended spatial
scale is summarized in Fig. \ref{Fgeo}. As argued above, transport processes
through the ECS and the astrocytic syncytium (panel A) can be described by a
simplified two domain model (panel B), which includes an ECS domain and an
astrocyte domain that exchange ions via typical astrocytic membrane mechanisms.
In this aspect, the astrocyte modelling framework is simpler than computational
neural models, which are often described with a high degree of spatial
specificity. However, in the spatial K\textsuperscript{+} buffering model, axial
fluxes in the ECS or inside the astrocyte must be described by the Nernst-Planck
equations for electrodiffusion (see Section~\ref{sec:math}). Then not only
electrical currents, but also ion concentration dynamics are explicitly
modelled. In this aspect, the astrocyte modelling framework is more complex than
computational neural models.

\subsection{Model constituents} \label{sec:math} 
In computational neuroscience
the main focus has been on modelling the short-term electrical signalling of
neurons. Integration of synaptic input and generation of action potentials
typically take place on a time scale of less than 100 milliseconds. At this
short time-scale, ionic concentrations of the main ionic charge carriers
(Na\textsuperscript{+}, K\textsuperscript{+} and Cl\textsuperscript{-}) vary
little. With the possible exception of the signalling molecule
Ca\textsuperscript{2+} (see e.g., \cite{Destexhe1996, Halnes2011}), ion
concentrations are typically assumed to be constant during the simulated period
or, at least, to vary so little that they do not evoke notable diffusive
currents. This is, for example, an underlying assumption in the cable equation
(e.g. \cite{Rall1977, Koch1999}), upon which most multicompartmental neural
models are built.

Contrarily, spatial K\textsuperscript{+} buffering by astrocytes typically takes
place at the time scale of seconds, and must account for ion concentration
changes. The dynamics of ion concentrations will not only depend on
transmembrane fluxes, but also on intra- and extracellular transports due to
diffusion (along concentration gradients) and electrical migration (along
voltage gradients) \cite{Chen2000, Halnes2013, Kofuji2004}. If ion concentration
gradients become sufficiently steep, the electrical currents associated with
diffusion of charged ions can be of comparable magnitude to the Ohmic currents
driven by electrical fields, and will therefore influence the electrodynamics of
the system \cite{Qian1989, Halnes2013}. An accurate description of long time
scale processing in brain tissue thus calls for an electrodiffusive formalism
based on the Nernst-Planck equations.

In earlier, pioneering modelling works on astrocytic
K\textsuperscript{+}-buffering \cite{Gardner-Medwin1983, Chen2000, Odette1988,
Newman1993}, the transient charge accumulation associated with the capacitive
membrane current was neglected. Due to this, the models might be inaccurate in
responses to transient signals, and do not guarantee a strictly consistent
relationship between electrical fields and ionic concentrations. Given that
K\textsuperscript{+}-buffering likely depends on an intricate balance between
diffusive and field-driven forces, this short-coming inspired us to develop a
new model \cite{Halnes2013}, which we present below. A key achievement of that
work was the derivation of the general mathematical equations for
electrodiffusion in a two domain setup as that in Fig. \ref{Fgeo}B.

\subsubsection{Kirchhoff-Nernst-Planck framework}
\label{sec:eldiff}
The challenge in solving the Nernst-Planck equations typically lies in how to compute the local electrical potential $v$. Generally, $v$ can be computed from Poisson's equations, as it is done in Poisson-Nernst-Planck solvers \cite{Leonetti1998, Lu2007, Lopreore2008, Nanninga2008, Pods2013}. However, to do this one must explicitly model charge relaxation processes, which demands an extremely fine spatiotemporal resolution \cite{Mori2009}. This makes Poisson-Nernst-Planck solvers computationally expensive and unsuited for predictions at a tissue level. For the K\textsuperscript{+}-buffering model \cite{Halnes2013}, we developed an alternative framework, which we may refer to as the Kirchhoff-Nernst-Planck (KNP) framework. The KNP-framework is summarized in Fig. \ref{Fgeo}C, and briefly introduced below. Useful entities and parameter definitions are listed in Table \ref{T2}. For a full derivation of this formalism, the reader is referred to the original work \cite{Halnes2013}.

The KNP-framework is a means of solving the continuity equations for the one-dimensional model system in Fig. \ref{Fgeo}B, for the intracellular
($n=I$) and ECS ($n=E$)  domains, for all ion species $k$ with valence $z_k$ and concentration $[k]_n$. The longitudinal intra- and extracellular fluxes are described by the Nernst-Planck equation, i.e., $j_{kn}$ in Fig. \ref{Fgeo}C, where first term on the right represents the diffusive flux density ($j_{kn}^{d}$), and the last term is the flux density due to ionic migration in the electrical field ($j_{kn}^{f}$).

Instead of using Poisson's equation to derive $v$, the KNP-framework derives the voltage gradients ($\partial v/\partial x$ in  Fig.~\ref{Fgeo}C) from the constraint that the sum of currents into a compartment should always be zero (i.e., Kirchhoff's current law). This zero-sum includes longitudinal diffusive- and field currents, transmembrane ionic currents and transmembrane capacitive currents. The three first of these are ionic currents, whereas the latter reflects the accumulation of charge on the astrocyte membrane. Hence, in the KNP-framework, the total electrical charge in a compartment is always consistent with the membrane charge associated with the membrane potential $v_M$, whereas the bulk solution is always electroneutral \cite{Halnes2013}. This is typically true for the bulk solution in brain tissue at time scales larger than 1 nanosecond~\cite{Grodzinsky2011}.

The KNP-framework has proven to be useful not only for modelling astrocytic buffering processes \cite{Halnes2013}, but also other processes taking place in brain tissue at long time scales \cite{Halnes2016, Halnes2017}. A three dimensional version of the KNP framework was recently developed \cite{Solbra2018}.

\subsubsection{Astrocyte model}
\label{sec:membmech}
The KNP-framework (Fig. \ref{Fgeo}C) is general to the choice of membrane mechanisms (reflected in the transmembrane flux density $j_{kM}$). In the modelling study in \cite{Halnes2013}, the astrocyte possessed the standard passive transmembrane fluxes of Na\textsuperscript{+} ($j_{Nap}$) and Cl\textsuperscript{-} ($j_{Clp}$), a passive K\textsuperscript{+}-flux ($j_{Kir}$) through the inward rectifying K\textsuperscript{+}-channel, and the fluxes through the Na\textsuperscript{+}/K\textsuperscript{+}-pump which exchanges 2 K\textsuperscript{+} (inward) for 3 Na\textsuperscript{+} ions (outward). The transmembrane fluxes of K\textsuperscript{+}, Na\textsuperscript{+} and Cl\textsuperscript{-} were, respectively
\begin{eqnarray}
j_{KM} = j_{Kir} - 2P
\label{jKm}
\\
j_{NaM}= j_{Nap} + 3P
\label{jNam}
\\
j_{ClM} = j_{Clp}.
\label{jClm}
\end{eqnarray}
The expressions for $j_{Kir}$, $j_{Nap}$, $j_{Clp}$ depend on the reversal potentials and conductances of the respective ion species, and the full expressions are given in the original work \cite{Halnes2013}. Also for the Na\textsuperscript{+}/K\textsuperscript{+}-pump rate, $P$, we refer to the original work for the full expression \cite{Halnes2013}.

\begin{table}[!ht]
\caption{\bf{Model parameters}}
\begin{tabular}{|l|r|l|}
  \hline
  \textbf{Parameter} & \textbf{Explanation} & \textbf{Value/Units} \\
  $k$ (index) & Ion species: $K^+$, $Na^+$ or $Cl^-$ & \\
  $n$ (index) & Domain: $I$ (ICS) or $E$ (ECS) &  \\
  $a_I$ & Astrocyte volume per total tissue volume  & 0.4 \\
  $a_E$ & ECS volume per total tissue volume & 0.2  \\
  $O_M$ & Astrocyte membrane area per total tissue volume  & m\textsuperscript{-1} \\
  $l$ & Length of astrocyte & $300 \, \mu \mathrm{m}$ \\
  $z_k$  & Valence of ion species $k$ & \\
  $[k]_n$ & Ion concentration of species $k$ in domain $n$ & mM \\
  $v_M$ & Membrane potential  & mV \\
  $j_{kM}$ & Membrane flux density of species $k$ & $\mu \mathrm{mol/(m}^2\mathrm{s})$ \\
  $j_{kn}^f$ & Axial flux density due to electrical migration & $\mu \mathrm{mol/(m}^2\mathrm{s})$ \\
  $j_{kn}^d$ & Axial flux density due to diffusion &  $\mu \mathrm{mol/(m}^2\mathrm{s})$ \\
  $j_{Kir}$ & Transmembrane $K^+$ flux density (Kir-channel) & $\mu \mathrm{mol/(m}^2\mathrm{s})$ \\
  $j_{Nap}$ & Passive, transmembrane $Na^+$ flux density  & $\mu \mathrm{mol/(m}^2\mathrm{s})$ \\
  $j_{Clp}$ & Passive, transmembrane $Cl^-$ flux density & $\mu \mathrm{mol/(m}^2\mathrm{s})$\\
  $P$ & Na\textsuperscript{+}/K\textsuperscript{+} exchanger pump-rate & $\mu \mathrm{mol/(m}^2\mathrm{s})$ \\
  $r_n$ & Resistivity &  $\Omega \mathrm{m}$\\
  $D_K$ & K\textsuperscript{+} diffusion constant & $1.96\times 10^{-9} \, \mathrm{m}^2/\mathrm{s}$ \\
  $D_{Na}$ & Na\textsuperscript{+} diffusion constant & $1.33\times 10^{-9} \, \mathrm{m}^2/\mathrm{s}$ \\
  $D_{Cl}$ & Cl\textsuperscript{-} diffusion constant & $2.03\times 10^{-9} \, \mathrm{m}^2/\mathrm{s}$  \\
  $\lambda_I$ & intracellular tortuosity & 3.2  \\
  $\lambda_E$ & extracellular tortuosity & 1.6  \\
  $C_{M}$ & specific membrane capacitance & $1 \, \mu \mathrm{F/cm}^2$\\
  $\Delta[k]_{n}$ & Deviance from baseline concentration & mM \\
  $^*[K^+]_{E}^{0}$ & baseline ECS K\textsuperscript{+}-concentration & $3.082 \, \mathrm{mM}$  \\
  $^*[K^+]_{I}^{0}$ & baseline ICS K\textsuperscript{+}-concentration & $99. 059 \, \mathrm{mM}$ \\
  $^*[Na^+]_{E}^{0}$ & baseline ECS Na\textsuperscript{+}-concentration & $144.622\, \mathrm{mM}$  \\
  $^*[Na^+]_{I}^{0}$ & baseline ICS Na\textsuperscript{+}-concentration & $15.189 \, \mathrm{mM}$  \\
  $^*[Cl^-]_{E}^{0}$ & baseline ECS Cl\textsuperscript{-}-concentration& $133.71 \, \mathrm{mM}$ \\
  $^*[Cl^-]_{I}^{0}$ & baseline ICS Cl\textsuperscript{-}-concentration & $5.145 \, \mathrm{mM}$ \\
  $^*v_{M0}$  & initial membrane potential & $-83.6 \, \mathrm{mV}$ \\
  $k_{dec}$ & decay factor for $[K^+]_E$ & $2.9\times 10^{-8} \, \mathrm{m/s}$ \\
  $j_{in}$ & constant input in input zone & $7\times 10^{-8} \, \mathrm{mol/(m}^2\mathrm{s})$\\
  \hline
  \end{tabular}
  \noindent
\begin{flushleft}
\textsuperscript{*} Initial values correspond to the resting state of the model (static steady state in the case of no external input). For the origin of the table values, see \cite{Halnes2013}.
\end{flushleft}
\label{T2}
\end{table}

\subsubsection{Simulation setup}
Below, we will present some simulations that explore how astrocytes transfer K\textsuperscript{+} out from high-concentration regions. To induce such a high-concentration region, a selected region of the ECS (the \emph{input zone} $0<x<l/10$ as indicated in Fig. \ref{Fgeo}B) was exposed to a constant influx ($j_{in}$) of K\textsuperscript{+} and a corresponding efflux of Na\textsuperscript{+}. In this way, the external input to the system was locally electroneutral, so that no net charge was added to the system (otherwise, the electroneutrality constraint would be violated). We assumed that this input signal reflected the effect of intense local AP firing in the input zone, where neurons take up Na\textsuperscript{+} and expel K\textsuperscript{+} from/to the ECS. Furthermore, we also implicitly included the effect of neuronal uptake mechanisms, i.e., neurons that take up K\textsuperscript{+} and expel Na\textsuperscript{+} via exchanger pumps. Unlike $j_{in}$, these mechanisms were evenly distributed in the system, so that all points in space $0<x<l$ experienced a K\textsuperscript{+}-concentration dependent decay towards the baseline ECS K\textsuperscript{+} concentration $[K^+]_{E}^0$. The input function was given by:
\begin{equation}
j_K^{in} = - j_{Na}^{in} = j_{in} -  k_{dec}([K^+]_E-[K^+]_E^0).
\label{input}
\end{equation}
Here, $j_{in}$ is a constant input which was applied only in the input zone ($0<x<l/10$) and in a selected time window
(100~s $< t <$ 400~s). The decay term was applied at all locations at all time. The decay factor ($k_{dec}$) and input flux density ($j_{in}$) are defined in Table ~\ref{T2}.

The model was implemented in Matlab, and the code is publicly available at ModelDB (https://senselab.med.yale.edu/ModelDB/ShowModel.cshtml?model=151945). We only included the dynamics of the main charge carriers (K\textsuperscript{+}, Na\textsuperscript{+} and Cl\textsuperscript{-}), and the simulations were run with sealed-end boundary conditions (see \cite{Halnes2013} for details).

\subsection{A mechanistic understanding of the K\textsuperscript{+}-buffering process}
\label{sec:flows}
Below, we aim to give a mechanistic picture of the K\textsuperscript{+}-buffering process by presenting selected simulations from the original study \cite{Halnes2013}. We focus on mapping out the transport routes of K\textsuperscript{+}, from entering to leaving the system, and on exploring how this depends on electrical forces, diffusive forces, and astrocytic uptake.

\subsection{Ion concentration dynamics and steady state}
\label{sec:SS}

\begin{figure}[!ht]
\begin{center}
\includegraphics[width=4in]{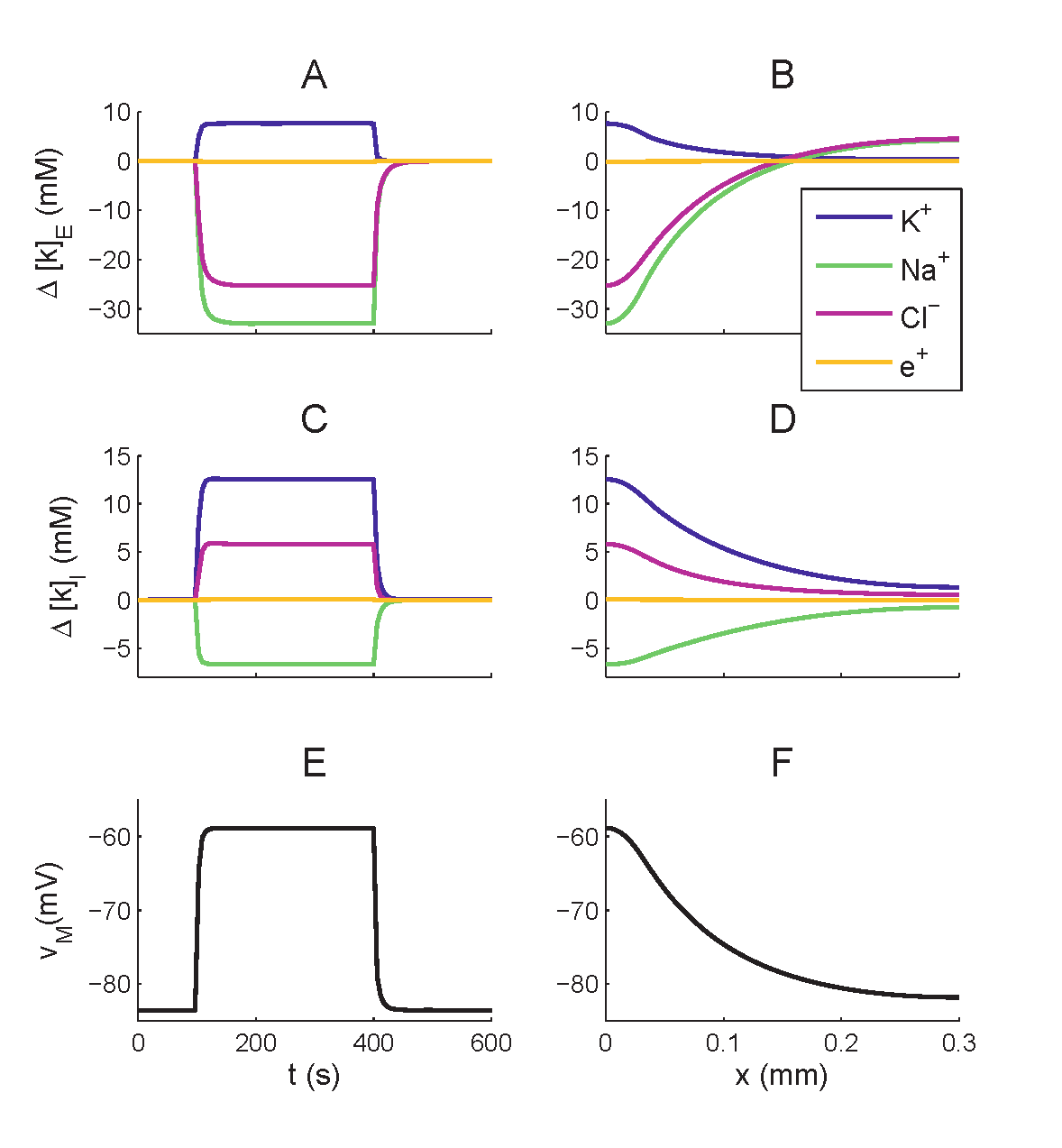}
\end{center}
\caption{
{\bf Dynamics and steady state profiles for the astrocyte/ECS-system.} (A, C, E) Dynamics of selected variables in a point ($x = 0$) in the input zone. The constant cation-exchange input was applied to the ECS of the input ($0<x<l/10$, $l$=0.3~mm) zone from $t=100$~s to $t=400$~s. During the input, ion concentrations in the ECS and ICS changed, but reached steady state after about 10-50 s after stimulus onset. $[K^+]_E$  (at $x=0$) had then increased by about 7.7 mM with respect to the baseline value (A), while $[K^+]_I$ had increased by about 12.5 mM due to uptake by the astrocyte (C). The astrocytic membrane potential had been depolarized to about -59 mV at $x=0$ (E). (B, D, F) Spatial profiles of selected variables at a time $t=400 \, \mathrm{s}$, when the system was in steady state. Deviations from the baseline ionic concentrations and $v_M$ were smaller outside the input zone, and typically decreased with $x$. Far away from the input zone ($x \sim l$=0.3~mm),
the conditions were close to the baseline conditions (A-D). Ionic concentrations were represented in terms of deviations from resting concentrations: $\Delta [k]_{n} = [k]_{n}-[k]_{n}^0$ for $n=I,E$. For direct comparison with ion concentrations, the charge density was represented as an equivalent concentration of unit charges $[e^+] = [K^+] + [Na^+] + [Cl^-]$. The figure was modified from \cite{Halnes2013}.
}
\label{Fprof}
\end{figure}

In Fig. \ref{Fprof} we see how the extracellular ion concentrations (panel A),
the intracellular ion concentrations (panel C), and $v_M$ (panel E) changed
after the input was turned on (at $t$=100~s). It took the system roughly 30
seconds to reach steady state, after which the state variables remained at fixed
values until the input was turned off (at $t$=400~s). For ion concentrations,
deviances from basal concentrations are shown. For example, $\Delta [K^{+}]_E$
was about 7.7 mM at steady state, corresponding to a concentration $[K^{+}]_E
\simeq 10.8 \, \mathrm{mM}$ (as the baseline concentration was $\sim 3.1 \,
\mathrm{mM}$). This value lies on the threshold between functional and
pathological conditions \cite{Chen2000, Newman1993, Hertz2013}, and is thus
likely to represent a scenario where the K\textsuperscript{+} buffering process
is of paramount importance. Although the input was applied to the ECS of the
input zone, $\Delta [K^{+}]_I \simeq 12.5 \, \mathrm{mM}$ was larger than
$\Delta [K^{+}]_E$, reflecting the astrocyte's propensity for local
K\textsuperscript{+}-uptake. As has been also seen experimentally
\cite{Newman1987, Dietzel1989, Chen2000}, the shifts in $[K^{+}]_E$ coincided
with a local depolarization of the astrocytic membrane (panel E).

In the following we focus on the spatial aspect of the buffering process and
limit the exploration to when the system is in steady state (the time $t$=400~s
was selected for the steady-state situation). For all system variables, the
deviation from the baseline conditions were biggest in the input zone, and
generally decayed with increasing $x$ (Fig.~\ref{Fprof}B,D,F). Note that the
properties of the system lead to a rearrangement of Cl\textsuperscript{-} in the
system, although Cl\textsuperscript{-} (unlike K\textsuperscript{+} and
Na\textsuperscript{+}) was not added/subtracted to/from the system. The
gradients of both the ionic concentrations and electrical potential were quite
pronounced, and we may thus expect that both diffusive and electrical forces
contribute to transporting ions through the system (from entering to leaving).
This is explored further in the following section.

\subsection{System throughput during steady state} 
\label{sec:flows} 
\begin{figure}[!h]
\begin{center}
\includegraphics[width=4in]{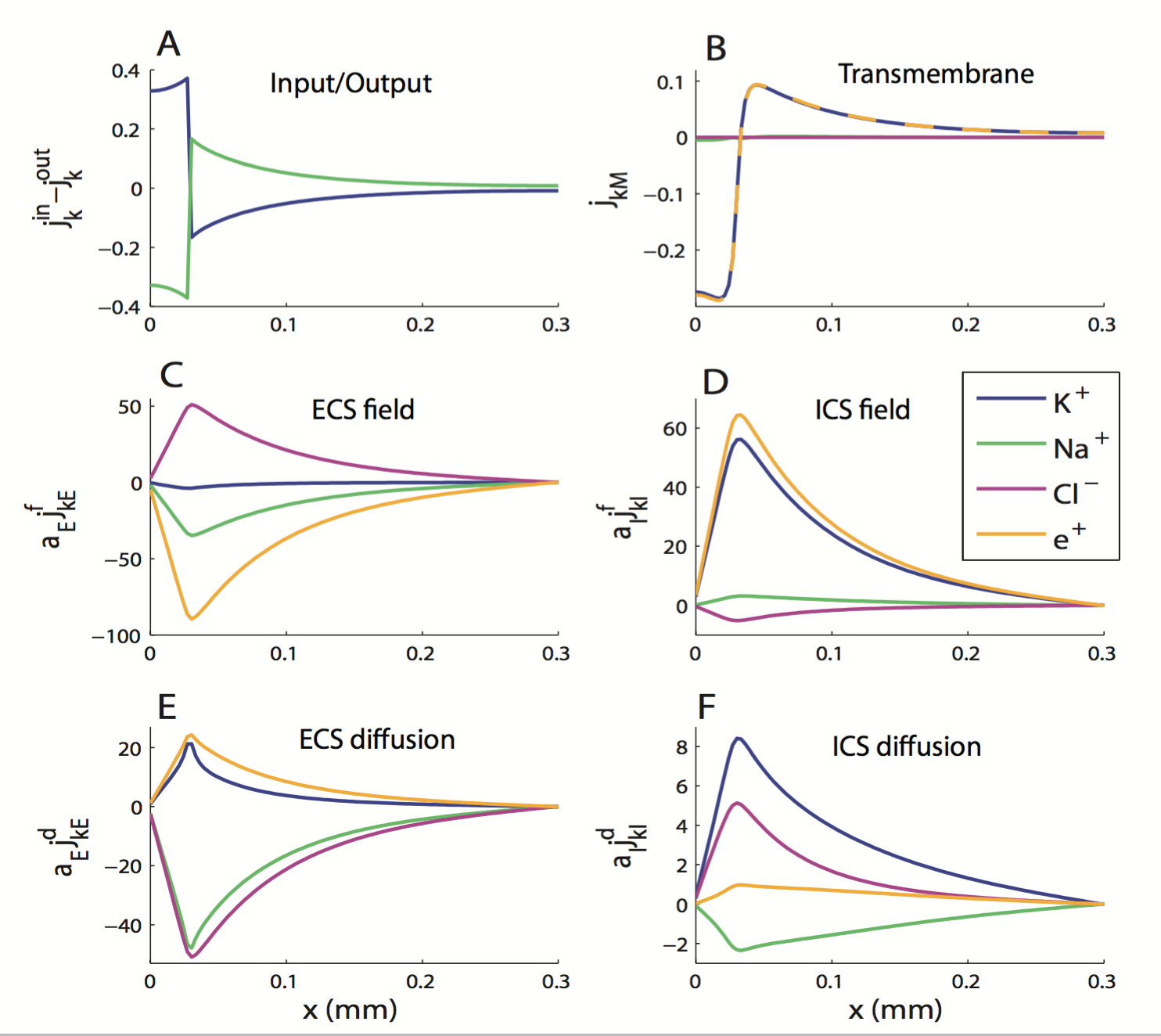} 
\end{center} 
\caption{ {\bf Ion transport in the astrocyte/ECS system at steady state.} 
(A) Total flux densities into system ($input-output$). (B) Transmembrane flux
densities. (C-F) Longitudinal flux densities due to (C) electrical migration in
the ECS, (D) electrical migration in the ICS, (E) diffusion in the ECS and (F)
diffusion in the ICS. (A-D) To aid comparison, flux densities $j_{kn}$ were
scaled by the relative area fraction $a_n$ (e.g., if $a_Ej_{kE} = a_Ij_{kI}$,
$I$ and $E$ carry the same the net flux of ion species $k$). For direct
comparison with ionic fluxes, the net electrical current was represented as a
flux of positive unit charges $j_{e+} = j_{K+} + j_{Na+} - j_{Cl-}$. 
The input zone was in the region $0<x<l/10$. Units on
the $y$-axis are $\mu \mathrm{mol/(m}^2\mathrm{s)}$ in all panels. The figure
was modified from \cite{Halnes2013} } 
\label{Fflows} 
\end{figure}

Figure \ref{Fflows} shows all the the ionic flows in the system during steady 
state. Due to the input configuration (Eq. \ref{input}) there was a net external influx
of K\textsuperscript{+} to the ECS in the input zone ($0<x<l/10$), and a net
efflux of K\textsuperscript{+} at all points outside this region
(Fig.~\ref{Fflows}A). As all external input/output was an electroneutral cation
exchange, Na\textsuperscript{+} and K\textsuperscript{+} had opposite input
profiles (compare blue and green curves).

\begin{figure}[!h]
\begin{center}
\includegraphics[width=4in]{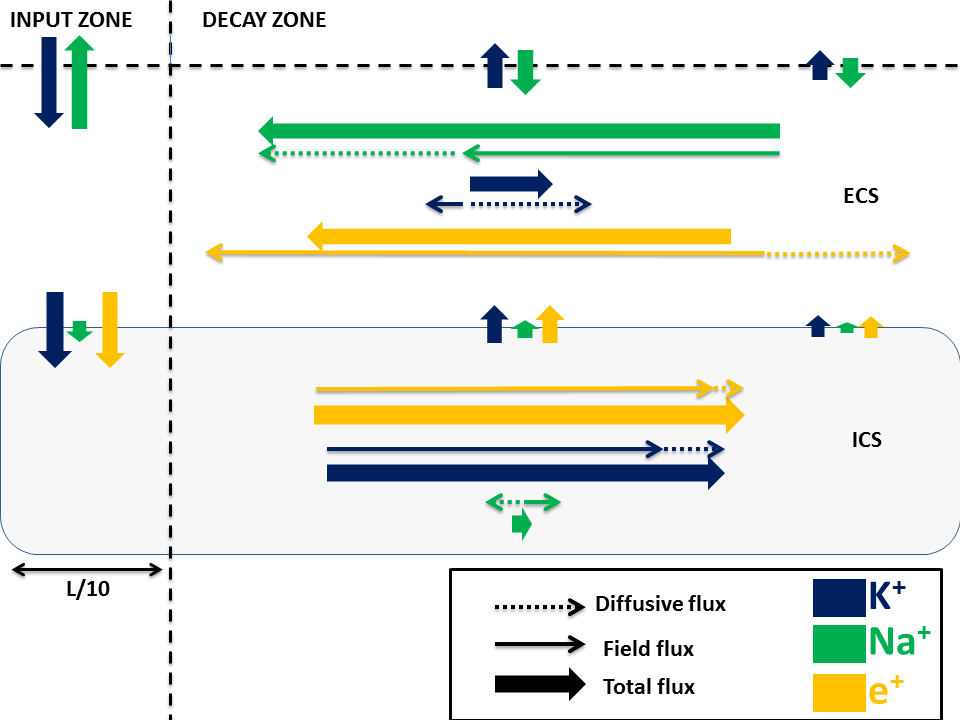} 
\end{center} 
\caption{ {\bf A flow chart of the ion transport in the astrocyte/ECS system at steady state.} 
A flow chart that qualitatively summarizes the essential information in Fig.\ref{Fflows}, showing
the main transport routes of K\textsuperscript{+} and Na\textsuperscript{+}
during steady state (Cl\textsuperscript{-} excluded from the overview).
K\textsuperscript{+} generally entered the system in the input zone and left the
system from some point along the astrocyte axis. The transport route of
K\textsuperscript{+} (from entering to leaving the system) was predominantly
intracellular, demonstrating the astrocyte's efficiency as a spatial buffer.
Na\textsuperscript{+} entered in the decay zone and left from the input zone.
Na\textsuperscript{+} transport predominantly took place in the ECS. The net
electrical current cycled in the system. The illustration is qualitative -
longer arrows mean higher flux densities, but the mapping from Fig.\ref{Fflows} 
to the flow chart is not quantitatively exact. The figure
was modified from \cite{Halnes2013} } 
\label{Fchart} 
\end{figure}

Our main interest was on how the astrocyte influences the route that
K\textsuperscript{+} follows through the system. Fig.~\ref{Fflows}B shows how
the transmembrane fluxes vary along the $x$-axis. The transmembrane flux of
K\textsuperscript{+} (dark blue) was much larger than for Na\textsuperscript{+}
(green) and Cl\textsuperscript{-} (magenta). By definition, the transmembrane
flux is negative for inward fluxes. According to Fig.~\ref{Fflows}B, there was
thus a pronounced astrocytic uptake of K\textsuperscript{+} in the
high-concentration region, and a release outside this region.

To complete the picture of the spatial buffering process, we must also look at
the intra- and extracellular fluxes of all ion species. To assess the relative
roles of diffusion and electrical migration, we distinguished between field
fluxes, which essentially reflect transports due to the extra-
(Fig.~\ref{Fflows}C) and intracellular (Fig.~\ref{Fflows}D) voltage gradients in
the system, and diffusive fluxes along the extra- (Fig.~\ref{Fflows}E) and
intracellular (Fig.~\ref{Fflows}F) concentration gradients in the system. As
expected, since the astrocyte absorbed K\textsuperscript{+} in the input zone
and released it outside this region (Fig.~\ref{Fflows}B), there was an
intracellular longitudinal transport of K\textsuperscript{+} in the positive
$x$-direction. This transport was partly due to diffusion (Fig.~\ref{Fflows}F),
as we could have predicted from the intracellular concentration gradient that we
saw in Fig. \ref{Fprof}D. However, due to the strong depolarization of the
astrocyte in the input zone (Fig.\ref{Fprof}F), electrical migration gave an
even larger contribution to the intracellular transport of K\textsuperscript{+}
(Fig.~\ref{Fflows}D).

K\textsuperscript{+} diffused in the positive $x$-direction also in the ECS
(Fig.~\ref{Fflows}E). However, a depolarization of the astrocyte corresponded to
a decrease in the ECS potential ($v_m = v_I-v_E$). Accordingly, the voltage
gradient in the ECS drove K\textsuperscript{+} in the negative $x$-direction
(Fig.~\ref{Fflows}C). That is, diffusion and electrical migration had
opposite directions in the ECS. This finding predicts that the astrocyte not only
provides an additional and more effective domain for longitudinal
K\textsuperscript{+}-transport, but even reduces the net transport of
K\textsuperscript{+} through the ECS, thus shielding the ECS from
K\textsuperscript{+}.

Figure \ref{Fchart} qualitatively summarizes the ionic flows through the system
during steady state. K\textsuperscript{+} entered the system in the ECS of the
input zone, where a major fraction of it crossed the membrane, was transported
intracellularly in the positive $x$-direction, and was released to the ECS at
higher $x$-values, from where it eventually left the system (via neuronal
uptake). Whereas the main transport route for K\textsuperscript{+} was
intracellular, the situation for Na\textsuperscript{+} was opposite.
Na\textsuperscript{+} entered the system outside the input zone, and was
predominantly transported in the negative $x$ direction through the ECS, and
left the system in the input zone (neuronal AP firing). The net
Cl\textsuperscript{-} transport was very small (flux densities due to diffusion
and electrical migration canceled each others out) and was not included in
Fig.\ref{Fchart}.

The simulations shown here illustrate that K\textsuperscript{+}-buffering
depends on a fine interplay between electrical and diffusive transport
processes, and could thus not be reliably simulated without a consistent
electrodiffusive modelling scheme, such as the KNP-framework. We conclude that
astrocytes seem tailored for shielding the ECS from excess K\textsuperscript{+}.
In the original work \cite{Halnes2013}, we also showed that K\textsuperscript{+}
clearance from high concentration was much more efficient when astrocytes were
present than in the case of a correspondingly enlarged ECS, a hypothesis that
had been posed earlier, but not yet tested \cite{Amedee1997}.

\section{Microscale water flow} \label{sec:microscale-flow}
A striking difference between astrocytes and neurons is that astrocytes have
water specific channels, aquaporin-4 (AQP4), embedded in the membrane. In the
astrocytic end-feet, i.e., the processes that reach out and collectively create
a sheath around the blood vessels and towards the pial surface, the AQP4 density
has been reported to be as high as $2400/\mathrm{\mu m}^2$ (see Furman et
al.~\cite{Furman:2003ik} Fig. 6). At the astrocytic membrane facing neuropil the
AQP4 density is about a factor 20 lower\cite{Enger:2012dy}, but this is still
dense compared to standard ion channel densities: in neurons the sodium channel
densities is estimated to be about $1/\mathrm{\mu m}^2$ in the cell body and
about $100/\mathrm{\mu m}^2$ to $200/\mathrm{\mu m}^2$ in the densest regions
such as the axon hillock \cite{Anonymous:d5AnHsy9}.

Throughout the body the lymphatic system removes excess fluid and waste from the tissue, but there is no
lymphatic system within neuropil, although lymph vessels were recently discovered
within the meningeal compartment \cite{Louveau:2015fua}. This lack of lymph
system within the neuropil led Iliff et al.~\cite{Iliff:2012jh} to suggest a
'glymphatic pathway' -- the brain's counterpart to a lymphatic pathway --
between the vasculature and the astrocyte end-feet, i.e., that a flow of water
through the interstitial space directed from the periarterial space to the
perivenous space (see Fig.~\ref{fig:osmotic_pressure_flow}E) functions as a
clearance mechanism for waste products such as amyloid beta. AQP4 deletion has
shown to reduce clearance of amyloid beta by $65\%$ \cite{Iliff:2012jh}, and
thus AQP4 seems to play an important role in the 'glymphatic system'.

The role of astrocytic AQP4 in potassium buffering is controversial~\cite{HajYasein:2015ge,Zhang:2008hg,AmiryMoghaddam:2003gq}. Here, we show how AQP4 can both stabilize the extracellular
concentration of solutes and possibly also induce a convective water flow. We
also show that there is a possible interplay between osmotic forces and water
fluxes through AQP4 interactions, which may be important for both potassium
buffering and for the glymphatic system.

\subsection{Transport mechanisms} 
The osmotic pressure over the membrane is
strong compared to typical hydrostatic pressures (see
Subsection~\ref{sec:Osmosis_and_astrocyte_swelling}). Upon osmotic changes,
water flows through the high density of AQP4 in the astrocyte membrane and
reestablishes the osmotic balance between the ECS and the intracellular volume
of the astrocytes. Depending on cortical
state, this may lead to swelling or shrinkage of the astrocytes. Xie et
al.~\cite{Xie:2013ck} found that during sleep the extracellular volume increases by about $60\%$ and that this increase in volume leads to a better clearance of amyloid beta
due to an increased convective flow through the ECS. Although osmosis may
regulate the ECS volume and thereby the convective flow, it is not clear whether
osmosis is a driving force for the flow itself. Cerebral arterial pulsation has,
however, been demonstrated to be a key driver for this flow \cite{Iliff:2013ja}.

Many aspects of the glymphatic system are not understood, and the concept itself
is still debated \cite{Hladky:2014ev,Smith:2015fh,Thrane:2015dk, Holter2017}. However, for water to be driven through the interstitial
space, from the periarterial to the perivenous space, either hydrostatic or
osmotic pressure gradients must act as driving forces. Note, however, that the
osmotic pressure and the hydrostatic pressure acts very differently: the
hydrostatic pressure may cause pressure difference between any two extracellular
(or intracellular) sites, whereas the osmotic pressure acts over membranes.

\begin{figure}[!htb] 
\centering
\includegraphics[width=3in]{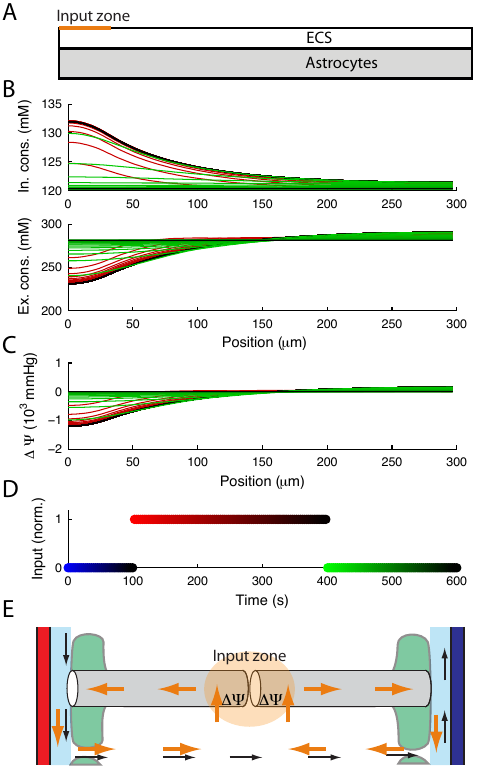}
\caption{{\bf Illustration of how the astrocyte machinery may generate an osmotic gradient following neural activity and induce water flow into the astrocyte.} The model shown in Fig.~\ref{Fgeo} is run with default parameters. (A) Illustration of the electrodiffusion model, see Fig.~\ref{Fgeo} for details. At time $t = 100~\mathrm{s}$ the model is triggered by neural activity (decreased extracellular Na\textsuperscript{+} and increased extracellular K\textsuperscript{+}) within the input zone. (B) The spatial distribution of net concentration of intracellular (upper) and extracellular ions (lower), leading to change in osmotic pressure seen in (C). The osmotic pressure is here assumed to be zero in equilibrium. (D) The color coding of (B) and (C) depicts time. Input is given in red, bright red at the beginning of the input, darker towards the end, whereas times after the $300~\mathrm{s}$ input stimulus is turned off is shown in green.
(E) An assumed system of two types of astrocytes, one type approaching the arterial site, the other type approaching the venous site. Red (left) is the arterial site, blue (right) is the venous site. The end-feet, shown in green, cover the vasculature and define perivascular spaces filled with water (light blue) along both arterioles and venules. Osmotically induced flow pattern depicted with orange arrows, and the hypothesized hydrostatic flow with black arrows. Note that the hydrostatic flow may go both through the end-feet and through the tiny clefts between end-feet.
}
\label{fig:osmotic_pressure_flow}
\end{figure}

If we assume an elevated hydrostatic pressure in the periarterial space compared
to the perivenous space, there will be a net hydrostatic pressure gradient over
the end-feet layers and through the interstitial space (see black arrows in
Fig.~\ref{fig:osmotic_pressure_flow}E). This may cause an extracellular flow,
and the convection may serve as a clearance mechanism for amyloid beta and other
waste products.

In Fig.~\ref{fig:osmotic_pressure_flow}, we have calculated pressure gradients
in the astrocyte/ECS model presented in Section~\ref{sec:Kbuff}, using identical
parameters and input conditions as there. The changes in ionic concentrations
induced by the input (cf. Eq.~\ref{input}), were most dramatic in the input
zone. In Fig.~\ref{fig:osmotic_pressure_flow}B, the summed intracellular and ECS
concentrations are shown at different time points after the input was turned on.
The concentration shifts gave rise to a maximal, activity-induced change in
osmolarity of more than $60~\mathrm{mOsm/L}$, i.e., the induced change in
concentrations between the intracellular and extracellular sites was more than
$60~\mathrm{mM}$. The pressure shift was largest in the input zone, where the
shift in osmotic pressure across the membrane was about $1000~\mathrm{mmHg}$
(see Fig.~\ref{fig:osmotic_pressure_flow}C, see also Eq.~\ref{eq:Morse} for the
formula for osmotic pressure). Further, the spatial extension of the cable lead
to an osmotic gradient along the membrane.

We did not explicitly include water flow in the modelling framework, but the
observations in Fig.~\ref{fig:osmotic_pressure_flow} allow us to make some
qualitative predictions from the pressure gradients. Firstly, from the osmotic
pressure in the input zone, we would predict a local influx of water, which
would lead to astrocyte swelling \cite{ostby_astrocytic_2009}. Secondly,
depending on the membrane elasticity, a hydrostatic pressure would build up
intracellularly, and cause an intracellular water flow away from the input zone.
Fig.~\ref{fig:osmotic_pressure_flow}E illustrates two different flow patterns
that we would expect if astrocytes were either connected to the arterial site
(left) or to the venous site (right). The high intracellular pressure would
cause an intracellular water flux out from the input zone (orange arrows),
directed both towards the arterial site and towards the venous site, depending
on astrocyte density, their connections to the vasculature, their respective
AQP4 densities et cetera. In Fig.~\ref{fig:osmotic_pressure_flow}E black arrows
indicate the expected flow caused by arterial pulsation and pressure differences
between the periarterial and perivenous spaces.

As of now, it is still an open question whether K\textsuperscript{+} buffering
evokes osmotic forces strong enough to produce a convective flow of considerable
magnitude. In general, it is not clear whether the net extracellular convective
flow is mainly driven by arterial pulsation or by osmotic forces. Experiments
show that both AQP4 and arterial pulsation are important for the glymphatic
flow: for AQP4 deletion clearance of amyloid beta was reduced by $65\%$
\cite{Iliff:2012jh}, and Iliff et al.~\cite{Iliff:2013ja} suggested that
cerebral arterial pulsatility is a key driver of paravascular CSF influx into
and through the brain parenchyma. It is presently not possible to measure the
small hydrostatic pressure differences within the tiny astrocytic compartments.
Thus, to understand the role of osmotic forces in potassium buffering a tight
interplay between experiments and modelling is needed. This also applies to the
glymphatic system and the question of clearance of waste products. Future
astrocyte models with spatial extension should emphasize to incorporate not only
electrodiffusion, but also the convective flow caused both by osmotic pressures
and by hydrostatic pressure originating from arterial pulsation or respiration.
Such a complete modelling framework is not available yet, but below we assess
some of the key ingredients that it will need.

\subsection{Osmosis and astrocyte swelling}
\label{sec:Osmosis_and_astrocyte_swelling} 
Above we showed some qualitative
results of possible convective flows, both extracellularly and intracellularly
in astrocytes. Here we will give quantitative estimates of the strength of the
osmotic forces, the water flow, and the typical time scale for this mechanism.

The lipid bilayer without AQP4 channels has a rather low water permeability
compared to the permeability of the astrocyte with a normal AQP4 density
\cite{Silberstein:2004ex}. As an additional boosting effect, AQP4 is often
arranged in so-called orthogonal arrays. These arrays may increase the
collective permeability of AQP4 as much as 10 times compared to the sum of the
single channel permeabilities \cite{Silberstein:2004ex}, and orthogonal arrays
are especially prominent within the astrocyte end-feet.

The membrane permeability, which is determined by the density of AQP4, has a
direct impact on how fast the ECS volume can be regulated upon osmotic stress.
In Tong et al.~\cite{Tong:2012jp} the ECS osmolarity was manipulated, and the
cells swelled until they reached a new equilibrium. The swelling was fitted to
exponential functions, and the estimated time constants were as small as
$\tau=1/(120.5~\mathrm{s}^{-1}) \approx 8.3~\mathrm{ms}$ for cells with high
densities of AQP4. In contrast, in \cite{Pangrsic2006} an astrocyte soma of diameter $\sim$ 10 $\mu$m was shown to swell with a time constant of tens of seconds, thereby stabilizing the ECS environment and presumably preventing osmotic pressure differences to cause swelling of neurons, as neurons do not have AQP4 water channels and thus have a much longer time constant for water inflow.

Below we describe the basic formalism for osmotically induced swelling of
astrocytes. Note that a similar formalism has been integrated with single
compartment models of ion dynamics 
(see e.g., ~\cite{ostby_astrocytic_2009, oyehaug_dependence_2011}), 
but not yet in spatially extended models.

\subsubsection{Basic formalism for osmotically induced volume changes} 
Similarly to Ohm's law for electricity, Darcy's law defines the water flow $Q$ as a
product of the permeability $G_\mathrm{tot}$ and the pressure difference $\Delta
\Psi$,
\begin{equation}
 Q=G_\mathrm{tot} \,\Delta \Psi~,
 \label{eq:Darcys}
\end{equation}
where $Q$ has units of $\mathrm{m^3/s}$, $\Delta \Psi$ has units of $\mathrm{Pa}$ and $G_\mathrm{tot}$ has units of $\mathrm{m^3/s/Pa}$. We use the symbol $\Psi$ for the water potential, since flow across a
membrane is caused both by hydrostatic and osmotic pressure differences over the
membrane,
\begin{equation}
  \Psi = \Psi_\mathrm{p}+\Psi_\Pi~.
\end{equation} 
Here $\Psi_\Pi$ is called the solute potential, and the hydrostatic pressure (the
pressure potential) is denoted $\Psi_\mathrm{p}$. Differences in the solute
potential over the membrane correspond to the osmotic pressure. Both $\Psi_\Pi$
and $\Psi_\mathrm{p}$ have units of Pascal. Water flows from a high water
potential to a lower water potential, and the solute water potential is defined
to be zero for pure water and is negative for any solute.

If we assume ideal solutions with low concentrations of solutes, we can use the
Morse equation to compute the solute potential \cite{Amiji2002}:
\begin{equation}
  \Psi_\Pi = i M R T~.
  \label{eq:Morse}
\end{equation}
Here $i$ is the ionization factor (van't Hoff factor), i.e., the ratio between
the actual concentration of particles produced when the substance is dissolved,
and the concentration of a substance as calculated from its mass, $R \approx
8.3~\mathrm{J}/(\mathrm{mol}\,\mathrm{K})$ is the gas constant, $T$, here
assumed to be $310~\mathrm{K}$, is the absolute temperature, and $M$ is the
osmotic concentration of solutes measured in moles per cubic meter. Our solutes are
here assumed to be ions, thus $i=1$.

If we combine Eqs.~\ref{eq:Darcys} and \ref{eq:Morse}, assume the hydrostatic
pressure $\Psi_\mathrm{p}$ to be the same at both sides of the membrane, and
notice that $Q$ is the flow rate of the extracellular volume $V_{e}$ across the
membrane and into the intracellular volume, i.e., $Q=\mathrm{d} V_{e}/\mathrm{d}
t$, the volume dynamics is given by
\begin{equation}
  \frac{\mathrm{d} V_\mathrm{e}}{\mathrm{d} t}=
  -RTG A \left(\frac{N_\mathrm{a}}{V-V_\mathrm{e}} -\frac{N_\mathrm{e}}{V_\mathrm{e}} \right)~.
\end{equation}
Here $N_\mathrm{e}$ and $N_\mathrm{a}$ are the number (moles) of extracellular
and intracellular (astrocytic) solutes, respectively, and
$V=V_\mathrm{e}+V_\mathrm{a}$ is the sum of the ECS and astrocyte volumes. We
have expressed the total membrane permeability $G_\mathrm{tot}$ from
Eq.~\ref{eq:Darcys} as $G_\mathrm{tot}=G A$, where $G$ is the permeability per
membrane area and $A$ is total membrane area within the volume $V$. If we assume
the volume $V$ and the number of ions $N_\mathrm{e}$ and $N_\mathrm{a}$ to be
constants, a general solution can be written
\begin{equation}
  -t RTGA=\frac{N_\mathrm{a} V V_\mathrm{e}}{(N_\mathrm{a}
  +N_\mathrm{e})^2}-\frac{V_\mathrm{e}^2}{2(N_\mathrm{a}+N_\mathrm{e})}
  + \frac{N_\mathrm{a} N_\mathrm{e} V^2 \ln ([N_\mathrm{a}+N_\mathrm{e}]V_\mathrm{e}
  -N_\mathrm{e}V)}{(N_\mathrm{a}+N_\mathrm{e})^3}+C~,
  \label{eq:osmosis_exponential}
\end{equation}
where 
$C$ is an integration constant. This expression cannot be converted into a simple formula 
for $V$ as a function of the time $t$, but for typical parameters 
(see Subsection~\ref{sec:experimental_parameters}), 
we find that the logarithmic term dominates the right hand side of the equality. 
Thus, an approximate solution can be written on a simple functional form as
\begin{equation}
  \frac{V_\mathrm{e}}{V} \approx \tilde{C} e^{-t/\tau}+\tilde{N}_\mathrm{e}~,
  \label{eq:volume_fraction_exponential}
\end{equation}
with the constant $\tilde{C}=V_0/V-\tilde{N}_\mathrm{e}$,
where $V_0$ is the initial extracellular volume and $\tilde{N}_\mathrm{e}=N_\mathrm{e}/(N_\mathrm{e}+N_\mathrm{a})$ is the relative extracellular number fraction. The time constant is given by
\begin{equation}
  \tau = \frac{N_\mathrm{a} N_\mathrm{e} V^2}{(N_\mathrm{e}+N_\mathrm{a})^3 R T G A}=\frac{\tilde{N}_\mathrm{a} \tilde{N}_\mathrm{e}}{c R T G A/V}~.
  \label{eq:tau_swell}
\end{equation}
In the last equality we have used the relative number fractions
$\tilde{N}_\mathrm{e}$ and
$\tilde{N}_\mathrm{a}=N_\mathrm{a}/(N_\mathrm{e}+N_\mathrm{a})$, and $c$ is the
net concentration of solutes within $V$, i.e.,
$c=(N_\mathrm{e}+N_\mathrm{a})/V$. From the expression we find that the time
constant is inversely proportional to the volume density $A/V$ of AQP4.

\subsubsection{Experimentally extracted parameters}
\label{sec:experimental_parameters}

\begin{table}[!ht]
  \begin{center}
  \caption{\bf{Model parameters, Section \ref{sec:microscale-flow}.}}
  \label{tab:osmotic_parameters}
\begin{tabular}{|l|r|l|} \hline
 \bf{Parameter} & \bf{Explanation} & \bf{Value/Units} \\ 
 $\tilde{N}_\mathrm{a}$ & fraction of intracellular ions & 0.55 \\ 
 $\tilde{N}_\mathrm{e}$ & fraction of extracellular ions & 0.45  \\ 
 $c$ & net solute concentration $c=(N_\mathrm{a}+N_\mathrm{e})/V$ & $300~\mathrm{mOsm/L}$  \\ 
 $R$ & gas constant & $8.314~\mathrm{LkPa/K/mol}$  \\ 
 $T$ & absolute temperature & $310~\mathrm{K}$  \\ 
 $P_\mathrm{f}$ & osmotic permeability per unit area & $0.01~\mathrm{cm/s}$ \\ 
 $G$ & water permeability per unit area & $7.0 \times 10^{-13}~\mathrm{m/s/Pa}$ \\ 
 $\Delta M$ & concentration difference of solutes & $60~\mathrm{mol/m^3}= 60~\mathrm{mM}$ \\ 
  $M_\mathrm{w}$ & molarity of water & $55.5~\mathrm{mol/L}= 55.5~\mathrm{M}$ \\ 
 $ Q_\mathrm{m}$ &  molar flux & $\mathrm{mol/s}$ \\
 $ Q $ &  transmembrane water flow & $\mathrm{m^3/s}$ \\
 $A/V$ & area to volume fraction & $20~\mathrm{\mu m^2/\mu m^3}$ \\ \hline
\end{tabular}
\end{center}
\end{table}

Based on experimentally measured quantities, we will in this subsection estimate typical the transmembrane water flow and the time constant $\tau$, derived in Eq.~\ref{eq:tau_swell}, for astrocytes. The most important parameters are found in Table~\ref{tab:osmotic_parameters}.

Estimates of the AQP4 water-channel permeability are typically based on osmotic
permeability experiments relating the concentration difference $\Delta M$ (in
units of $\mathrm{mol/m^3}$) of impermeable molecules between two sides of a
cell membrane, to the molar flux $Q_\mathrm{m}$ of water through the membrane,
\begin{equation}
  P_\mathrm{f,cell}=Q_\mathrm{m}/\Delta M~.
  \label{eq:P_f}
\end{equation}
The molar flux $Q_\mathrm{m}$ can be estimated from
cell swelling measurements. 
Further, the single channel AQP4 permeability or the permeability per
unit area can be estimated by additional electron microscopy of the density of
AQP4 channels within the membrane \cite{Tong:2012jp}. If we assume the
hydrostatic pressure $\Psi_\mathrm{p}$ to be the same at both sides of the
membrane, the correspondence between reported osmotic permeability per unit area
$P_\mathrm{f}$ and the permeability per unit area $G=G_\mathrm{tot}/A$ where $A$
is the membrane area, is found by substituting Eqs.~\ref{eq:Morse} and
\ref{eq:P_f} into Eq.~\ref{eq:Darcys}. This gives
\begin{equation}
  G = P_\mathrm{f}/(R T M_\mathrm{w})~,
\end{equation}
with $M_\mathrm{w}=Q_\mathrm{m}/Q$ denoting the molarity of water.

The permeability per unit area $G$ was used in Eq.~\ref{eq:osmosis_exponential}
to relate the volumetric flow rate $Q=\mathrm{d} V_\mathrm{e}/\mathrm{d} t$
through the membrane to the osmotic pressure difference, and it was shown that
its value is inversely proportional to the time constant $\tau$,
see~Eq.~\ref{eq:tau_swell}. Based on data for the relative density of AQP4 in
end-feet vs.~the membrane facing neuropil \cite{Enger:2012dy}, as
well as estimates for both $P_\mathrm{f}$ \cite{Silberstein:2004ex, Jung:1994tv}
and single channel permeability \cite{Tong:2012jp, Hashido:2007fr}, we use
$P_\mathrm{f}=0.01~\mathrm{cm/s}$ to arrive at a numerical value for the time
constant $\tau$.

The total neuronal membrane area is estimated to be $25000~\mathrm{m^2}$ for
human \cite[p.~97]{Bear:2015vb} and the brain volume is about
$1350~\mathrm{cm^3}$. This leads to a membrane area to volume density of
$18.5~\mathrm{\mu m^2/\mu m^3}$. A similar density can be found for mouse. We
assume a default value for the volume density of astrocytic membrane area to be
similar to the volume density of neuronal membrane area, i.e.,
$A/V=20~\mathrm{\mu m^2/\mu m^3}$.

We further assume increased osmotic pressure equivalent to $1000~\mathrm{mmHg}$
within a cube of sides $100~\mathrm{\mu m}$. According to our assumptions this
volume will contain $A_\mathrm{m}=2 \cdot 10^7~\mathrm{\mu m^2}$ of membrane
area. The osmotic pressure corresponds to a concentration difference of less
than $60~\mathrm{mOsm/L}$ between the inside and the outside of the membrane, cf.
Eq.~\ref{eq:Morse}. The net rate of water influx to the astrocytes within the
volume is then
\begin{equation}
  Q = G A_\mathrm{m} \Delta \Psi = 0.002 \mathrm{mm^3/s}~.
\end{equation}
This initial influx rate corresponds to twice the total volume $V$ per second,
and for a standard extracellular volume fraction of $20~\%$ it corresponds to a
net influx rate of 10 times the ECS volume per second. However, as water moves
from the ECS to the intracellular space, the osmotic pressure will be reduced
according to Eq.~\ref{eq:osmosis_exponential}. If we assume both the ECS and the
astrocytic volume fraction to be $20\%$ of the total volume \cite{Hertz:2006jw},
the astrocytic osmolarity to be $330~\mathrm{mOsm/L}$, and the extracellular
osmolarity to be $270~\mathrm{mOsm/L}$, the system will, according to
Eqs.~\ref{eq:volume_fraction_exponential} and \ref{eq:tau_swell}, reach
equilibrium when the extracellular volume fraction is $18\%$ and the
intracellular is $22\%$, and the time constant will be $\tau=9.2~\mathrm{ms}$.
This is in good agreement with experimental results from Tong et
al.~\cite{Tong:2012jp}, and is a consequence of the high area to volume density of astrocytic membrane area within the brain.

These simple considerations show that the osmotic response to neural activity is
fast and the transmembrane influx rate of water may be substantial within a given volume of tissue, although single astrocyte somas swell substantially slower when located in an extracellular solution with constant extracellular osmolarity \cite{Pangrsic2006}. 
 
\section{Macroscale fluid flow through brain tissue} \label{sec:macroscale-flow}
Fluid flow through the central nervous system is inherently of a
multiscale character and thus amenable to a multitude of modelling
approaches, for a recent review in the context of biomechanics see,
e.g.,~\cite{Goriely2015}. At the macroscale, that is, at the whole
brain scale, three distinctive approaches to the continuum modelling
of fluid and solute flow through the brain tissue are: (i) single or
multiple-network poroelasticity theory, (ii) direct modeling and
coupling of the interstitial tissue flow with free flow (Darcy-Stokes
flow)~\cite{BaberEtAl2012}, and (iii) coupling the microcirculation
(vasculature) with the tissue itself via one-dimensional networks
(Darcy flow with embedded Navier-Stokes
flow)~\cite{CattaneoZunino2014}. Of these, the first approach has
gained momentum for the biomechanical modelling of fluid flow through
brain tissue and also through the spinal cord, see
e.g.,~\cite{Goriely2015} and references therein.

These single- and multiple-network poroelasticity theories originally
originate from geoscience~\cite{Biot1941, BaiElsworthRoegiers1993},
but have been proposed to model the flow and exchange of fluid through
biological tissue in general and the brain in
particular~\cite{TullyVentikos2011}. The theory is based on modeling
the tissue as a solid matrix permeated by interconnecting multiple
pathways or networks where each network is characterized by its
porosity. For instance, Vardakis et
al~\cite{VardakisTullyVentikos2013} consider a model incorporating the
intracellular space permeated by four networks corresponding to the
arteries, the arterioles/capillaries, the veins and the extracellular
space respectively. In the case of a single fluid network (here, the
extracellular space) and under the assumptions of linear tissue
elasticity, isotropy, and incompressibility, the general
multiple-network poroelasticity model reduces to the well-known Biot
equations~\cite{Biot1941}. In addition, to modelling the flow of
interstitial fluid through the tissue itself, computational models
typically also need to account for the interplay with the
cerebrospinal fluid flow in the subarachnoid spaces and ventricles as
illustrated by, e.g.,~\cite{Rutkowska2012, Stoverud2013,
  VardakisTullyVentikos2013}.

The multiple-network poroelasticity equations, in the absence of local
body forces, sources or sinks, and inertial terms, and assuming
incompressible fluid networks for the sake of simpler presentation,
take the form~\cite{TullyVentikos2011}: find the spatially and
temporally varying tissue deformation $u$ and the pressures $p_{a}$
for each network $a = 1, \dots, A$ such that
\begin{equation}
  \label{eq:mpet:momentum}
  - \nabla \cdot \sigma(u) + \sum_{a} \nabla p_{a} = 0,
\end{equation}
where $\sigma$ is the stress tensor, for instance given by Hooke's law
under the assumptions of linearly deformating elastic tissue, and for
each $a = 1, \dots, A$:
\begin{equation}
  \label{eq:mpet:mass}
  \nabla \cdot \dot{u}
  - \nabla \cdot ( G_{a} \nabla p_{a} )
  = \sum_{b \not = a} s_{b \rightarrow a}.
\end{equation}
In~Eq.~\ref{eq:mpet:mass}, $G_{a} = \kappa_{a}/\mu_{a}$ where
$\kappa_{a}$ is the permeability and $\mu_{a}$ the viscosity of
network $a$. The exchange coefficients $s_{b \rightarrow a}$ govern
the rate of transfer from network $b$ to network $a$ and are subject
to modelling. The fluid velocity in network $a$ is defined in terms of
the network pressure $p_a$ by:
\begin{equation}
  \label{eq:fluid-velocity}
  v_{a} = - G_{a} \phi_{a}^{-1} \nabla p_{a},
\end{equation}
where $\phi_{a}$ is the porosity or volume fraction of network $a$
which, note, may or may not be spatially varying.

The multiple-network poroelasticity framework is attractive in that it
allows for specific incarnations, at different levels of complexity,
and modelling fluid flow in, e.g., the different compartments of the
'glymphatic system'. Variations in the extracellular and astrocytic
volume fractions, cf.~Section
\ref{sec:Osmosis_and_astrocyte_swelling}, are naturally modelled via
the network porosities, while microscale water flow and AQP4
densities, as discussed in Section~\ref{sec:microscale-flow}, govern the
dynamics of the exchange coefficients.

\section{Concluding remarks}

The recent years have seen the launching of several ambitious large-scale
projects aiming to link the different scales in the brain, from molecules to
systems. Prominent examples are EUs Human Brain Project, the US BRAIN Initiative
and Project MindScope at Allen Institute of Brain Sciences~\cite{Kandel2013}.
Here computational neuroscience will play a key role in integrating
neurobiological data, storing current knowledge, and testing hypotheses by means
of a set of interconnected mathematical models together bridging the different
scales. The main focus of these projects is to understand the information
processing in the brain, and a central goal is to link processing and spiking
activity at the single-neuron level to systems level measures of neural activity
such as the local field potential (LFP)~\cite{Einevoll2013},
electroencephalograpy (EEG)~\cite{Nunez2006}, or magnetoencephalography
(MEG)~\cite{Hamalainen1993}.

In these projects the astrocytes will be secondary to the neurons as their main
role is likely to be of a homeostatic nature, i.e., to provide the neurons with
stable environments for their information processing (but see other chapters in
the present volume). An example is the presently discussed role of astrocytes in
providing spatial K$^+$ buffering to maintain low concentrations of potassium in
the ECS and avoid pathological neuronal firing.

Another multiscale brain project is to try to link neural activity to
macroscopic fluid flow. Here astrocytes will likely be at center stage as their
ion-concentration dynamics appear to provide a key connection to fluid flow via
osmotic effects. In the present chapter we have outlined some elements that we
believe will be of key importance for this multiscale approach: In
Section~\ref{sec:pathology} we discussed how neuronal and astrocytic dynamics
are linked via ions in ECS, in Section~\ref{sec:Kbuff} we described the recently
introduced Kirchoff-Nernst-Planck (KNP) scheme for modeling ion dynamics in
astrocytes (and brain tissue in general), in Section~\ref{sec:microscale-flow}
we described how astrocytes may regulate microscopic liquid flow by osmotic
effects, and in Section~\ref{sec:macroscale-flow} we finally discussed how such
microscopic flow can be linked to whole-brain macroscopic flow.

A long-term ultimate goal would be to make joint multiscale, multimodal models
for brain tissue including both information processing by spikes, liquid flow,
hemodynamic activity and metabolic activity~\cite{Devor2012}. Such models could
not only become invaluable mathematical "microscopes" for exploring brain
function but also hopefully become important tools for understanding brain
pathologies and suggest new treatments.

\section{Funding and acknowledgments}
G. Halnes was supported by the Research Council of Norway (NFR), through Digital Life project "Digibrain" 248828. M. E. Rognes was supported by the European Research Council (ERC) under the European Union's Horizon 2020 research and innovation programme under grant agreement 714892.

\begin{small}
\bibliography{Kbuff}

\begin{thebibliography}{10}
\providecommand{\url}[1]{\texttt{#1}}
\providecommand{\urlprefix}{URL }
\expandafter\ifx\csname urlstyle\endcsname\relax
  \providecommand{\doi}[1]{doi:\discretionary{}{}{}#1}\else
  \providecommand{\doi}{doi:\discretionary{}{}{}\begingroup
  \urlstyle{rm}\Url}\fi
\providecommand{\bibAnnoteFile}[1]{%
  \IfFileExists{#1}{\begin{quotation}\noindent\textsc{Key:} #1\\
  \textsc{Annotation:}\ \input{#1}\end{quotation}}{}}
\providecommand{\bibAnnote}[2]{%
  \begin{quotation}\noindent\textsc{Key:} #1\\
  \textsc{Annotation:}\ #2\end{quotation}}
\providecommand{\eprint}[2][]{\url{#2}}

\bibitem{Wang2008}
Wang DD, Bordey A (2008) {The astrocyte odyssey.}
\newblock Progress in neurobiology 86: 342--67.
\bibAnnoteFile{Wang2008}

\bibitem{somjen_ions_2004}
Somjen GG (2004) Ions in the Brain: Normal Function, Seizures, and Stroke.
\newblock Oxford University Press, {USA}, 1 edition.
\bibAnnoteFile{somjen_ions_2004}

\bibitem{frankenhaeuser_after-effects_1956}
Frankenhaeuser B, Hodgkin AL (1956) The after-effects of impulses in the giant
  nerve fibres of loligo.
\newblock J Physiol 131: 341--76.
\bibAnnoteFile{frankenhaeuser_after-effects_1956}

\bibitem{Cordingley1978}
Cordingley G, Somjen G (1978) {The clearing of excess potassium from
  extracellular space in spinal cord and cerebral cortex}.
\newblock Brain research 151: 291--306.
\bibAnnoteFile{Cordingley1978}

\bibitem{Dietzel1989}
Dietzel I, Heinemann U, Lux H (1989) {Relations between slow extracellular
  potential changes, glial potassium buffering, and electrolyte and cellular
  volume changes during neuronal hyperactivity in cat}.
\newblock Glia 2: 25--44.
\bibAnnoteFile{Dietzel1989}

\bibitem{Gardner-Medwin1983}
Gardner-Medwin A (1983) {Analysis of potassium dynamics in mammalian brain
  tissue.}
\newblock The Journal of physiology : 393--426.
\bibAnnoteFile{Gardner-Medwin1983}

\bibitem{Chen2000}
Chen KC, Nicholson C (2000) {Spatial buffering of potassium ions in brain
  extracellular space.}
\newblock Biophysical journal 78: 2776--97.
\bibAnnoteFile{Chen2000}

\bibitem{Haj-Yasein2014}
Haj-Yasein NN, Bugge CE, Jensen V, Ostby I, Ottersen OP, et~al. (2014)
  {Deletion of aquaporin-4 increases extracellular K(+) concentration during
  synaptic stimulation in mouse hippocampus.}
\newblock Brain structure \& function 220: 2469--74.
\bibAnnoteFile{Haj-Yasein2014}

\bibitem{Hertz2013}
Hertz L, Xu J, Song D, Yan E, Gu L, et~al. (2013) {Astrocytic and neuronal
  accumulation of elevated extracellular K+ with a 2/3 K+/Na+ flux
  ratio—consequences for energy metabolism, osmolarity and higher brain
  function}.
\newblock Frontiers in Computational Neuroscience 7: 1--22.
\bibAnnoteFile{Hertz2013}

\bibitem{Newman1993}
Newman EA (1993) {Inward-rectifying potassium channels in retinal glial
  (M\"{u}ller) cells.}
\newblock The Journal of neuroscience : the official journal of the Society for
  Neuroscience 13: 3333--45.
\bibAnnoteFile{Newman1993}

\bibitem{Sykova2008}
Sykov\'{a} E, Nicholson C (2008) {Diffusion in Brain Extracellular Space}.
\newblock Physiol Rev 88: 1277--1340.
\bibAnnoteFile{Sykova2008}

\bibitem{Enger2015}
Enger R, Tang W, Vindedal GF, Jensen V, {Johannes Helm} P, et~al. (2015)
  {Dynamics of Ionic Shifts in Cortical Spreading Depression.}
\newblock Cerebral cortex (New York, NY : 1991) : 1--8.
\bibAnnoteFile{Enger2015}

\bibitem{Park2006}
Park EH, Durand DM (2006) {Role of potassium lateral diffusion in non-synaptic
  epilepsy: a computational study.}
\newblock Journal of theoretical biology 238: 666--82.
\bibAnnoteFile{Park2006}

\bibitem{florence_role_2009}
Florence G, Dahlem MA, Almeida ACG, Bassani JWM, Kurths J (2009) The role of
  extracellular potassium dynamics in the different stages of ictal bursting
  and spreading depression: a computational study.
\newblock J Theor Biol 258: 219--228.
\bibAnnoteFile{florence_role_2009}

\bibitem{somjen2001mechanisms}
Somjen GG (2001) Mechanisms of spreading depression and hypoxic spreading
  depression-like depolarization.
\newblock Physiological reviews 81: 1065--1096.
\bibAnnoteFile{somjen2001mechanisms}

\bibitem{Orkand1966}
Orkand RK, Nicholls JG, Kuffler SW (1966) {Effect of nerve impulses on the
  membrane potential of glial cells in the central nervous system of amphibia.}
\newblock Journal of neurophysiology 29: 788--806.
\bibAnnoteFile{Orkand1966}

\bibitem{lux_ionic_1986}
Lux HD, Heinemann U, Dietzel I (1986) Ionic changes and alterations in the size
  of the extracellular space during epileptic activity.
\newblock Adv Neurol 44: 619--39.
\bibAnnoteFile{lux_ionic_1986}

\bibitem{oyehaug_dependence_2011}
{\O}yehaug L, {\O}stby I, Lloyd CM, Omholt SW, Einevoll GT (2012) Dependence of
  spontaneous neuronal firing and depolarisation block on astroglial membrane
  transport mechanisms.
\newblock J Comput Neurosci 32: 147--165.
\bibAnnoteFile{oyehaug_dependence_2011}

\bibitem{de2008dysfunctional}
De~Keyser J, Mostert JP, Koch MW (2008) Dysfunctional astrocytes as key players
  in the pathogenesis of central nervous system disorders.
\newblock Journal of the neurological sciences 267: 3--16.
\bibAnnoteFile{de2008dysfunctional}

\bibitem{Nedergaard2005}
Nedergaard M, Dirnagl U (2005) {Role of glial cells in cerebral ischemia.}
\newblock Glia 50: 281--6.
\bibAnnoteFile{Nedergaard2005}

\bibitem{Kriv1975}
Kr\'{\i}z N, Sykov\'{a} E, Vyklick\'{y} L (1975) {Extracellular potassium
  changes in the spinal cord of the cat and their relation to slow potentials,
  active transport and impulse transmission.}
\newblock The Journal of physiology 1: 167--182.
\bibAnnoteFile{Kriv1975}

\bibitem{Lothman1975}
Lothman E, Somjen G (1975) {Extracellular potassium activity, intracellular and
  extracellular potential responses in the spinal cord.}
\newblock The Journal of physiology 1: 115--136.
\bibAnnoteFile{Lothman1975}

\bibitem{Coles1986}
Coles J, Orkand R (1986) {Free Concentrations of Na, K, and Cl in the Retina of
  the Honeybee Drone: Stimulus‐Induced Redistribution and Homeostasisa}.
\newblock Annals of the New \ldots 481: 303--317.
\bibAnnoteFile{Coles1986}

\bibitem{Nicholson2000}
Nicholson C, Chen K, Hrab\v{e}tov\'{a} S, Tao L (2000) {Diffusion of molecules
  in brain extracellular space: theory and experiment}.
\newblock Progress in brain research 125: 129--154.
\bibAnnoteFile{Nicholson2000}

\bibitem{Halnes2013}
Halnes G, Ostby I, Pettersen KH, Omholt SW, Einevoll GT (2013)
  {Electrodiffusive model for astrocytic and neuronal ion concentration
  dynamics.}
\newblock PLoS computational biology 9: e1003386.
\bibAnnoteFile{Halnes2013}

\bibitem{Macaulay2012}
Macaulay N, Zeuthen T (2012) {Glial K⁺ clearance and cell swelling: key roles
  for cotransporters and pumps.}
\newblock Neurochemical research 37: 2299--309.
\bibAnnoteFile{Macaulay2012}

\bibitem{Kofuji2004}
Kofuji P, Newman EA (2004) {Potassium buffering in the central nervous system.}
\newblock Neuroscience 129: 1045--56.
\bibAnnoteFile{Kofuji2004}

\bibitem{ostby_astrocytic_2009}
{\O}stby I, {\O}yehaug L, Einevoll GT, Nagelhus EA, Plahte E, et~al. (2009)
  Astrocytic mechanisms explaining neural-activity-induced shrinkage of
  extraneuronal space.
\newblock {PLoS} Comp Biol 5: e1000272.
\bibAnnoteFile{ostby_astrocytic_2009}

\bibitem{ullah_models_2009}
Ullah G, Schiff S (2009) Models of epilepsy.
\newblock Scholarpedia 4: 1409.
\bibAnnoteFile{ullah_models_2009}

\bibitem{hubel_dynamics_2014}
H{\"u}bel N, Dahlem MA (2014) Dynamics from {Seconds} to {Hours} in
  {Hodgkin}-{Huxley} {Model} with {Time}-{Dependent} {Ion} {Concentrations} and
  {Buffer} {Reservoirs}.
\newblock PLoS Comput Biol 10: e1003941.
\bibAnnoteFile{hubel_dynamics_2014}

\bibitem{kager_simulated_2000}
Kager H, Wadman WJ, Somjen GG (2000) Simulated seizures and spreading
  depression in a neuron model incorporating interstitial space and ion
  concentrations.
\newblock J Neurophysiol 84: 495--512.
\bibAnnoteFile{kager_simulated_2000}

\bibitem{kager_seizure-like_2006}
Kager H, Wadman WJ, Somjen GG (2006) Seizure-like afterdischarges simulated in
  a model neuron.
\newblock J Comput Neurosci 22: 105--128.
\bibAnnoteFile{kager_seizure-like_2006}

\bibitem{cressman_influence_2009}
Cressman J, Ullah G, Ziburkus J, Schiff S, Barreto E (2009) The influence of
  sodium and potassium dynamics on excitability, seizures, and the stability of
  persistent states: I. single neuron dynamics.
\newblock J Comput Neurosci .
\bibAnnoteFile{cressman_influence_2009}

\bibitem{Somjen2008}
Somjen GG, Kager H, Wadman WJ (2008) {Computer simulations of neuron-glia
  interactions mediated by ion flux.}
\newblock Journal of computational neuroscience 25: 349--65.
\bibAnnoteFile{Somjen2008}

\bibitem{sibille2015neuroglial}
Sibille J, Duc KD, Holcman D, Rouach N (2015) The neuroglial potassium cycle
  during neurotransmission: role of kir4. 1 channels.
\newblock PLoS Comput Biol 11: e1004137.
\bibAnnoteFile{sibille2015neuroglial}

\bibitem{jensen_role_1997}
Jensen MS, Yaari Y (1997) Role of intrinsic burst firing, potassium
  accumulation, and electrical coupling in the elevated potassium model of
  hippocampal epilepsy.
\newblock J Neurophysiol 77: 1224--33.
\bibAnnoteFile{jensen_role_1997}

\bibitem{ziburkus_interneuron_2006}
Ziburkus J, Cressman JR, Barreto E, Schiff SJ (2006) Interneuron and pyramidal
  cell interplay during in vitro seizure-like events.
\newblock J Neurophysiol 95: 3948--54.
\bibAnnoteFile{ziburkus_interneuron_2006}

\bibitem{grisar_contribution_1992}
Grisar T, Guillaume D, Delgado-Escueta AV (1992) Contribution of
  {Na}+,{K}(+)-{ATPase} to focal epilepsy: a brief review.
\newblock Epilepsy Res 12: 141--149.
\bibAnnoteFile{grisar_contribution_1992}

\bibitem{verkhratsky2013glial}
Verkhratsky A, Butt AM (2013) Glial physiology and pathophysiology.
\newblock John Wiley \& Sons.
\bibAnnoteFile{verkhratsky2013glial}

\bibitem{zhang2010astrocyte}
Zhang Y, Barres BA (2010) Astrocyte heterogeneity: an underappreciated topic in
  neurobiology.
\newblock Current opinion in neurobiology 20: 588--594.
\bibAnnoteFile{zhang2010astrocyte}

\bibitem{Destexhe1996}
Destexhe A, Bal T, McCormick DA, Sejnowski TJ, Sejnowski J, et~al. (1996)
  {Ionic mechanisms underlying synchronized oscillations and propagating waves
  in a model of ferret thalamic slices.}
\newblock J Neurophysiol 76: 2049--2070.
\bibAnnoteFile{Destexhe1996}

\bibitem{Halnes2011}
Halnes G, Augustinaite S, Heggelund P, Einevoll GT, Migliore M (2011) {A
  multi-compartment model for interneurons in the dorsal lateral geniculate
  nucleus.}
\newblock PLoS Comput Biol 7: e1002160.
\bibAnnoteFile{Halnes2011}

\bibitem{Rall1977}
Rall W (1977) {Core conductor theory and cable properties of neurons}.
\newblock In: Kandel E, Brookhardt J, {Mountcastle VM}, editors, Handbook of
  Physiology, Bethesda: American Physiological Society, chapter~3. pp. 39--97.
\newblock
  \urlprefix\url{http://onlinelibrary.wiley.com/doi/10.1002/cphy.cp010103/full}.
\bibAnnoteFile{Rall1977}

\bibitem{Koch1999}
Koch C (1999) {Biophysics of computation: information processing in single
  neurons.}
\newblock Oxford University Press: New York, 1st edition.
\bibAnnoteFile{Koch1999}

\bibitem{Qian1989}
Qian N, Sejnowski T (1989) {An electro-diffusion model for computing membrane
  potentials and ionic concentrations in branching dendrites, spines and
  axons}.
\newblock Biological Cybernetics 15: 1--15.
\bibAnnoteFile{Qian1989}

\bibitem{Odette1988}
Odette L, Newman EA (1988) {Model of potassium dynamics in the central nervous
  system}.
\newblock Glia 210: 198--210.
\bibAnnoteFile{Odette1988}

\bibitem{Leonetti1998}
L\'{e}onetti M, Dubois-Violette E (1998) {Theory of Electrodynamic
  Instabilities in Biological Cells}.
\newblock Physical Review Letters 81: 1977--1980.
\bibAnnoteFile{Leonetti1998}

\bibitem{Lu2007}
Lu B, Zhou YC, Huber Ga, Bond SD, Holst MJ, et~al. (2007) {Electrodiffusion: a
  continuum modeling framework for biomolecular systems with realistic
  spatiotemporal resolution.}
\newblock The Journal of chemical physics 127: 135102.
\bibAnnoteFile{Lu2007}

\bibitem{Lopreore2008}
Lopreore CL, Bartol TM, Coggan JS, Keller DX, Sosinsky GE, et~al. (2008)
  {Computational modeling of three-dimensional electrodiffusion in biological
  systems: application to the node of Ranvier.}
\newblock Biophysical journal 95: 2624--35.
\bibAnnoteFile{Lopreore2008}

\bibitem{Nanninga2008}
Nanninga PM (2008) {A computational neuron model based on Poisson – Nernst
  – Planck theory}.
\newblock ANZIAM J 50: 46--59.
\bibAnnoteFile{Nanninga2008}

\bibitem{Pods2013}
Pods J, Sch\"{o}nke J, Bastian P (2013) {Electrodiffusion models of neurons and
  extracellular space using the Poisson-Nernst-Planck equations--numerical
  simulation of the intra- and extracellular potential for an axon model.}
\newblock Biophysical journal 105: 242--54.
\bibAnnoteFile{Pods2013}

\bibitem{Mori2009}
Mori Y (2009) {From Three-Dimensional Electrophysiology to the Cable Model : an
  Asymptotic Study}.
\newblock arXiv:09013914 [q-bioNC] : 1--39.
\bibAnnoteFile{Mori2009}

\bibitem{Grodzinsky2011}
Grodzinsky F (2011) {Fields, Forces, and Flows in Biological Systems.}
\newblock Garland Science, Taylor \& Francis Group, London \& New York.
\bibAnnoteFile{Grodzinsky2011}

\bibitem{Halnes2016}
Halnes G, M{\"{a}}ki-Marttunen T, Keller D, Pettersen KH, Andreassen OA, et~al.
  (2016) {Effect of Ionic Diffusion on Extracellular Potentials in Neural
  Tissue}.
\newblock PLOS Computational Biology 12: e1005193.
\bibAnnoteFile{Halnes2016}

\bibitem{Halnes2017}
Halnes G, M{\"{a}}ki-Marttunen T, Pettersen KH, Andreassen OA, Einevoll GT
  (2017) {Ion diffusion may introduce spurious current sources in
  current-source density (CSD) analysis}.
\newblock Journal of Neurophysiology 118: 114--120.
\bibAnnoteFile{Halnes2017}

\bibitem{Solbra2018}
Solbr{\aa} A, Bergersen AW, van~den Brink J, Malthe-S{\o}renssen A, Einevoll
  GT, et~al. (2018) A kirchhoff-nernst-planck framework for modeling large
  scale extracellular electrodiffusion surrounding morphologically detailed
  neurons.
\newblock PLOS Computational Biology 14: 1-26.
\bibAnnoteFile{Solbra2018}

\bibitem{Newman1987}
Newman EA (1987) {Distribution of potassium conductance in mammalian Muller
  (glial) cells: a comparative study}.
\newblock The Journal of neuroscience 7: 2423--2432.
\bibAnnoteFile{Newman1987}

\bibitem{Amedee1997}
Amedee T, Robert A, Coles J (1997) {Potassium homeostasis and glial energy
  metabolism}.
\newblock Glia : 599--630.
\bibAnnoteFile{Amedee1997}

\bibitem{Furman:2003ik}
Furman CS, Gorelick-Feldman DA, Davidson KGV, Yasumura T, Neely JD, et~al.
  (2003) {Aquaporin-4 square array assembly: opposing actions of M1 and M23
  isoforms.}
\newblock Proceedings of the National Academy of Sciences of the United States
  of America 100: 13609--13614.
\bibAnnoteFile{Furman:2003ik}

\bibitem{Enger:2012dy}
Enger R, Gundersen GA, Haj-Yasein NN, Eilert-Olsen M, Thoren AE, et~al. (2012)
  {Molecular scaffolds underpinning macroglial polarization: an analysis of
  retinal M{\"u}ller cells and brain astrocytes in mouse.}
\newblock Glia 60: 2018--2026.
\bibAnnoteFile{Enger:2012dy}

\bibitem{Anonymous:d5AnHsy9}
Safronov BV, Wolff M, Vogel W (1999) {Axonal expression of sodium channels in
  rat spinal neurones during postnatal development.}
\newblock The Journal of Physiology 514 ( Pt 3): 729--734.
\bibAnnoteFile{Anonymous:d5AnHsy9}

\bibitem{Louveau:2015fua}
Louveau A, Smirnov I, Keyes TJ, Eccles JD, Rouhani SJ, et~al. (2015)
  {Structural and functional features of central nervous system lymphatic
  vessels.}
\newblock Nature 523: 337--341.
\bibAnnoteFile{Louveau:2015fua}

\bibitem{Iliff:2012jh}
Iliff JJ, Wang M, Liao Y, Plogg BA, Peng W, et~al. (2012) {A Paravascular
  Pathway Facilitates CSF Flow Through the Brain Parenchyma and the Clearance
  of Interstitial Solutes, Including Amyloid $\beta$}.
\newblock Science Translational Medicine 4: 147ra111--147ra111.
\bibAnnoteFile{Iliff:2012jh}

\bibitem{HajYasein:2015ge}
Haj-Yasein NN, Bugge CE, Jensen V, {\O}stby I, Ottersen OP, et~al. (2015)
  {Deletion of aquaporin-4 increases extracellular K(+) concentration during
  synaptic stimulation in mouse hippocampus.}
\newblock Brain structure {\&} function 220: 2469--2474.
\bibAnnoteFile{HajYasein:2015ge}

\bibitem{Zhang:2008hg}
Zhang H, Verkman AS (2008) {Aquaporin-4 independent Kir4.1 K+ channel function
  in brain glial cells.}
\newblock Molecular and cellular neurosciences 37: 1--10.
\bibAnnoteFile{Zhang:2008hg}

\bibitem{AmiryMoghaddam:2003gq}
Amiry-Moghaddam M, Williamson A, Palomba M, Eid T, de~Lanerolle NC, et~al.
  (2003) {Delayed K+ clearance associated with aquaporin-4 mislocalization:
  phenotypic defects in brains of alpha-syntrophin-null mice.}
\newblock Proceedings of the National Academy of Sciences of the United States
  of America 100: 13615--13620.
\bibAnnoteFile{AmiryMoghaddam:2003gq}

\bibitem{Xie:2013ck}
Xie L, Kang H, Xu Q, Chen MJ, Liao Y, et~al. (2013) {Sleep Drives Metabolite
  Clearance from the Adult Brain}.
\newblock Science 342: 373--377.
\bibAnnoteFile{Xie:2013ck}

\bibitem{Iliff:2013ja}
Iliff JJ, Wang M, Zeppenfeld DM, Venkataraman A, Plog BA, et~al. (2013)
  {Cerebral arterial pulsation drives paravascular CSF-interstitial fluid
  exchange in the murine brain.}
\newblock Journal of Neuroscience 33: 18190--18199.
\bibAnnoteFile{Iliff:2013ja}

\bibitem{Hladky:2014ev}
Hladky SB, Barrand MA (2014) {Mechanisms of fluid movement into, through and
  out of the brain: evaluation of the evidence.}
\newblock Fluids and Barriers of the CNS 11: 26.
\bibAnnoteFile{Hladky:2014ev}

\bibitem{Smith:2015fh}
Smith AJ, Jin BJ, Verkman AS (2015) {Muddying the water in brain edema?}
\newblock Trends in neurosciences 38: 331--332.
\bibAnnoteFile{Smith:2015fh}

\bibitem{Thrane:2015dk}
Thrane AS, Rangroo~Thrane V, Plog BA, Nedergaard M (2015) {Filtering the
  muddied waters of brain edema.}
\newblock Trends in neurosciences 38: 333--335.
\bibAnnoteFile{Thrane:2015dk}

\bibitem{Holter2017}
Holter KE, Kehlet B, Devor A, Sejnowski TJ, Dale AM, et~al. (2017)
  {Interstitial solute transport in 3D reconstructed neuropil occurs by
  diffusion rather than bulk flow}.
\newblock Proceedings of the National Academy of Sciences : 201706942.
\bibAnnoteFile{Holter2017}

\bibitem{Silberstein:2004ex}
Silberstein C (2004) {Membrane organization and function of M1 and M23 isoforms
  of aquaporin-4 in epithelial cells}.
\newblock AJP: Renal Physiology 287: F501--F511.
\bibAnnoteFile{Silberstein:2004ex}

\bibitem{Tong:2012jp}
Tong J, Briggs MM, McIntosh TJ (2012) {Water Permeability of Aquaporin-4
  Channel Depends on BilayerComposition, Thickness, and Elasticity}.
\newblock Biophysical Journal 103: 1899--1908.
\bibAnnoteFile{Tong:2012jp}

\bibitem{Pangrsic2006}
Pangrsic T, Potokar M, Haydon PG, Zorec R, Kreft M (2006) {Astrocyte swelling
  leads to membrane unfolding, not membrane insertion.}
\newblock Journal of neurochemistry 99: 514--23.
\bibAnnoteFile{Pangrsic2006}

\bibitem{Amiji2002}
Amiji MM, Sandmann BJ (2002) Applied Physical Pharmacy.
\newblock McGraw-Hill.
\bibAnnoteFile{Amiji2002}

\bibitem{Jung:1994tv}
Jung JS, Bhat RV, Preston GM, Guggino WB, Baraban JM, et~al. (1994) {Molecular
  characterization of an aquaporin cDNA from brain: candidate osmoreceptor and
  regulator of water balance.}
\newblock Proceedings of the National Academy of Sciences of the United States
  of America 91: 13052--13056.
\bibAnnoteFile{Jung:1994tv}

\bibitem{Hashido:2007fr}
Hashido M, Kidera A, Ikeguchi M (2007) {Water transport in aquaporins: osmotic
  permeability matrix analysis of molecular dynamics simulations.}
\newblock Biophysical Journal 93: 373--385.
\bibAnnoteFile{Hashido:2007fr}

\bibitem{Bear:2015vb}
Bear MF, Paradiso MA, Connors BW (2001) {Neuroscience: Exploring the Brain}.
\newblock Baltimore: Lippincott Williams and Wilkins, 2nd edition.
\bibAnnoteFile{Bear:2015vb}

\bibitem{Hertz:2006jw}
Hertz L, Peng L, Dienel GA (2006) {Energy metabolism in astrocytes: high rate
  of oxidative metabolism and spatiotemporal dependence on
  glycolysis/glycogenolysis}.
\newblock Journal of cerebral blood flow and metabolism : official journal of
  the International Society of Cerebral Blood Flow and Metabolism 27: 219--249.
\bibAnnoteFile{Hertz:2006jw}

\bibitem{Goriely2015}
Goriely A, Geers MG, Holzapfel GA, Jayamohan J, Jérusalem A, et~al. (2015)
  Mechanics of the brain: perspectives, challenges, and opportunities.
\newblock Biomechanics and Modeling in Mechanobiology : 1-35.
\bibAnnoteFile{Goriely2015}

\bibitem{BaberEtAl2012}
Baber K, Mosthaf K, Flemisch B, Helmig R, M{\"u}thing S, et~al. (2012)
  Numerical scheme for coupling two-phase compositional porous-media flow and
  one-phase compositional free flow.
\newblock IMA journal of applied mathematics 77: 887--909.
\bibAnnoteFile{BaberEtAl2012}

\bibitem{CattaneoZunino2014}
Cattaneo L, Zunino P (2013) Computational models for coupling tissue perfusion
  and microcirculation.
\newblock MOX Report 25/2013 .
\bibAnnoteFile{CattaneoZunino2014}

\bibitem{Biot1941}
Biot MA (1941) General theory of three-dimensional consolidation.
\newblock Journal of applied physics 12: 155--164.
\bibAnnoteFile{Biot1941}

\bibitem{BaiElsworthRoegiers1993}
Bai M, Elsworth D, Roegiers JC (1993) Multiporosity/multipermeability approach
  to the simulation of naturally fractured reservoirs.
\newblock Water Resources Research 29: 1621--1633.
\bibAnnoteFile{BaiElsworthRoegiers1993}

\bibitem{TullyVentikos2011}
Tully BJ, Ventikos Y (2011) Cerebral water transport using multiple-network
  poroelastic theory: application to normal pressure hydrocephalus.
\newblock Journal of Fluid Mechanics 667: 188--215.
\bibAnnoteFile{TullyVentikos2011}

\bibitem{VardakisTullyVentikos2013}
Vardakis JC, Tully BJ, Ventikos Y (2013) Exploring the efficacy of endoscopic
  ventriculostomy for hydrocephalus treatment via a multicompartmental
  poroelastic model of {CSF} transport: A computational perspective.
\newblock PloS ONE 8: 1--16.
\bibAnnoteFile{VardakisTullyVentikos2013}

\bibitem{Rutkowska2012}
Rutkowska G, Haughton V, Linge S, Mardal KA (2012) Patient-specific 3d
  simulation of cyclic csf flow at the craniocervical region.
\newblock American Journal of Neuroradiology 33: 1756--1762.
\bibAnnoteFile{Rutkowska2012}

\bibitem{Stoverud2013}
St{\o}verud K, Langtangen H, Haughton V, Mardal K (2013) Csf pressure and
  velocity in obstructions of the subarachnoid spaces.
\newblock The neuroradiology journal 26: 218--226.
\bibAnnoteFile{Stoverud2013}

\bibitem{Kandel2013}
Kandel ER, Markram H, Matthews PM, Yuste R, Koch C (2013) Neuroscience thinks
  big (and collaboratively).
\newblock Nat Rev Neurosci 14: 659--664.
\bibAnnoteFile{Kandel2013}

\bibitem{Einevoll2013}
Einevoll G, Kayser C, Logothetis N, Panzeri S (2013) Modelling and analysis of
  local field potentials for studying the function of cortical circuits.
\newblock Nature Reviews Neuroscience 14: 770--785.
\bibAnnoteFile{Einevoll2013}

\bibitem{Nunez2006}
Nunez PL, Srinivasan R (2006) Electric fields of the brain: The Neurophysics of
  EEG.
\newblock Oxford University Press, Inc., 2nd ed. edition.
\bibAnnoteFile{Nunez2006}

\bibitem{Hamalainen1993}
H{\"a}m{\"a}l{\"a}inen M, Hari R, Ilmoniemi R, Knuutila J, Lounasmaa OV (1993)
  Magnetoencephalography - theory, instrumentation, and applications to
  noninvasive studies of the working human brain.
\newblock Reviews of Modern Physics 65: 413--497.
\bibAnnoteFile{Hamalainen1993}

\bibitem{Devor2012}
Devor A, Boas D, Einevoll G, Buxton R, Dale A (2012) Neuronal basis of
  non-invasive functional imaging: from bold fmri to microscopic neurovascular
  dynamics.
\newblock In: Choi IY, Gruetter R, editors, Neural Metabolism In Vivo,
  Springer. pp. 433-500.
\bibAnnoteFile{Devor2012}

\end{thebibliography}
\end{small}

\end{document}